\documentclass[10pt,preprint]{aastex}

\usepackage{graphicx,natbib}

\shorttitle{Orbital motion of HR 8799b, c, and d using HST data from 1998}
\shortauthors{Soummer et al.}

\begin{document}
\title{Orbital Motion of HR 8799 b,c, d using Hubble Space Telescope data from 1998: Constraints on Inclination, Eccentricity and Stability}

\slugcomment{Accepted for publication in ApJ August 25, 2011}

\author{R\'emi Soummer, J. Brendan Hagan, Laurent Pueyo\altaffilmark{1}, Adrien Thormann\altaffilmark{2}, Abhijith Rajan, Christian Marois\altaffilmark{3}}
\affil{Space Telescope Science Institute, 3700 San Martin Drive, Baltimore MD 21218, USA}
\email{email: soummer@stsci.edu}




\altaffiltext{1}{Department of Physics and Astronomy, Johns Hopkins University, Baltimore, MD, USA}
\altaffiltext{2}{Department of Mechanical Engineering, Johns Hopkins University, Baltimore, MD, USA}

\altaffiltext{3}{NRC Herzberg Institute of Astrophysics, Victoria, Canada}
\marginparwidth = 10pt
\hoffset = 0pt
\begin{abstract}

 HR 8799 is currently the only multiple-planet system that has been detected with direct imaging, with four giant planets of masses $7-10$ $M_{Jup}$ orbiting at large separations (15-68 AU) from this young late A star. Orbital motion provides insight into the stability, and possible formation mechanisms of this planetary system. Dynamical studies can also provide constraints on the planets' masses, which help calibrate evolutionary models. Yet, measuring the orbital motion is a very difficult task because the long-period orbits (50-500 yr) require long time baselines and high-precision astrometry.
This paper studies the three planets HR 8799b, c and d in the archival data set of HR 8799 obtained with the Hubble Space Telescope (HST) NICMOS coronagraph in 1998.
The detection of all three planets is made possible by a careful optimization of the LOCI algorithm, and we used a statistical analysis of a large number of reduced images.
This work confirms previous astrometry for planet b, and presents new detections and astrometry for c and d.
These HST images provide a ten-year baseline with the discovery images from 2008, and therefore offer a unique opportunity to constrain their orbital motion now.
Recent dynamical studies of this system show the existence of a few possible stable solutions involving mean motion resonances, where the interaction between c and d plays a major role. We study the compatibility of a few of these stable scenarios (1d:1c, 1d:2c, or 1d:2c:4d) with the new astrometric data from HST. In the hypothesis of a 1d:2c:4b mean motion resonance our best orbit fit is close to the stable solution previously identified for a three-planet system, and involves low eccentricity for planet d ($e_d=0.10$) and moderate inclination of the system ($i=28.0$ deg), assuming a coplanar system, circular orbits for b and c, and exact resonance with integer period ratios. Under these assumptions, we can place strong constraints on the inclination of the system ($27.3-31.4$ deg) and on the eccentricity for d $e_d<  0.46$.  Our results are robust to small departures from exact integer period ratios, and consistent with previously published results based on dynamical studies for a three-planet system prior to the discovery of the fourth planet.

\end{abstract}


\maketitle 
\section{Introduction}\label{sect:intro}

Direct imaging of exoplanets is a particularly difficult problem because of the very small angular separation and very high-contrast between the planet and the star \citep{OH09}. The first images of planets orbiting nearby stars have been obtained relatively recently \citep{MMB08,KGC08,LGC09,LBC10,MZK10}
 and followed up at different wavelengths \citep{QMK10,CTM11,BLB11}. In addition, other interesting substellar companions to stars and brown dwarfs have also been discovered using direct imaging \citep{NOK95,CLD05,TCJ09,CBF10,LJV10,BLW10}. Several surveys have already been conducted to search for faint companions \citep{LDM07,KAJ07,NCB08,LSH10,ELM10}, and direct imaging will expand in the near future with new facility instruments on large ground-based telescopes \citep{MGP08,BFD08,HST08,HOZ11}.

 It is interesting to note that multiple systems have been discovered by every major observing techniques \citep{WUM09,GN10}. These include discoveries using radial velocities \citep{FMB08,RLB10}, transits \citep{HFR10,LFF11,LRF11}, microlensing \citep{GBU08}, and pulsar timing \citep{Wol94}. HR 8799 is currently the only multiple system discovered using direct imaging.   The three planets b, c and d were discovered by \citet{MMB08}. The fourth planet was discovered by \citet{MZK10}, and confirmed with a single epoch image at VLT by \citet{CBI11}. Some of these HR 8799 planets have also been imaged in archival data sets using other observatories \citep{FIT09,LMD09,MMZ09,MMV10}. Because the system's discovery is recent there are still few data points available to constrain the general configuration of the system (e.g. planet masses, inclination, eccentricities, periods). This translates into a limited understanding of the formation scenarios, dynamical stability of the system, and the atmospheric properties. In this paper, we provide new astrometric and photometric measurements with a ten year baseline that will help progress in each of these areas.

HR 8799 is a $\lambda$ Bootis star with spectral type similar to A5V and F0V, and is located at 39.4 pc \citep{GK99}. Age is a critical parameter for understanding direct images, since the observed near-infrared radiation originates from the gravitational potential energy that was released and converted into heat during their formation. As young giant planets cool down with time, they become less luminous and harder to detect, but the cooling process lasts hundreds of millions of years \citep{MFH07}.  With an age estimate between 30 and 60 MYr \citep{MZK10,ZRS11}, the HR 8799 planets are relatively bright and at relatively large angular separations ($\sim$ 0.4 to $\sim$1.7 arcsec) which makes them excellent targets for direct imaging. 

The four giant planets orbit at large separations ( 15-68 AU), and the system also exhibits a complex disk architecture including a warm asteroid belt analog within 6-15AU, a cold Kuiper belt analog between 90-300 AU, and an extended halo of small grains up to $\sim1000$ AU \citep{SRS09}. Estimates for the inclination of the system using different methods are consistent with a mostly face-on system with all four planets orbiting in the same direction (counter clock-wise). Inclination estimates are typically below 30-40 degrees \citep{LMD09,SRS09,MRS10}, with dynamical stability in favor of an inclination larger than 20 degrees \citep{RKS09}. The star is likely to be significantly more inclined (40 degrees) than the system \citep{WCD11}.

Such a multiple-planet system is of remarkable interest, with all four planets likely to have formed in the same protoplanetary disk, with same age and metallicity. 
The presence of these massive objects at large separations makes this system a particularly interesting target from a planetary formation standpoint. 
 The formation mechanisms for these planets either using gravitational instability or core accretion is an active topic of research \citep{Raf05,KMY10,Boss11,CBI11}.

The dynamical stability of the system over time periods of the order of 100 Myr places upper limits on the planet masses, since stability favors lower masses in a tightly packed multi-planet system. 
These mass constraints also help validate evolutionary and atmospheric models in this mass regime, which still lack comparison with real data. 
A lower limit for the masses can be obtained from evolutionary models for the minimum age of the system, since the star is on the main sequence.
Currently the consensus mass ranges are: 6-7 $M_{Jup}$ for b, and 7-10 $M_{Jup}$ for c, d, and e \citep{MMB08,MZK10,CBI11}. 

In the absence of astrometric measurements spanning a significant portion of the orbits, dynamical and stability studies have so far focused on integrating a number of possible scenarios (e.g. coplanar circular orbits, non coplanar orbits, eccentric orbits) based on initial conditions derived from the discovery data from 2004 and 2008. The few stable solutions identified by these studies are stabilized by mean motion resonances (MMR) mechanisms \citep{GM09,RKS09,FM10,MHC10}. 
Orbital characterization using a long temporal baseline is a key element for such stability studies, as it provides a possibility to test the compatibility of the scenarios with the data. 
The HST data from 1998 offers a unique opportunity to constrain today the astrometry of this planetary system with a ten year baseline, and therefore brings insight into orbital configuration, dynamical stability with possible resonances, and implications for planet masses and formation scenarios.

Understanding the planet's atmospheric properties also poses a number of challenges. 
Their SEDs can be fitted with an L-dwarf style model, but the difficulty is to fit a temperature low enough to give a radius large enough to be consistent with brown dwarf cooling tracks (e.g. \citet{MMB08}).
Models using thick cloud layers that are not present in brown dwarfs tend to reproduce the planets' SEDs with higher fidelity \citep{MMB08,CBI11,BMK11,MBC11}.  The focus of this paper is on astrometry, and the analysis of the new photometric information for planet c and d in the F160W filter is beyond the scope of this paper.

The 1998 images of HR 8799 in the HST data set, were obtained with the NICMOS coronagraph using the 0.63 arcsec occulting spot. 
Advanced point spread function (PSF) subtractions based on roll-deconvolution or angular differential imaging \citep{SS03,MLD06,LMD07} are required because the coronagraph alone is not sufficient to provide sufficient dynamic range, and the images are dominated by residual quasi-static speckles that limit the detection sensitivity \citep{AS04,SFA07,HOS07}. The HST/NICMOS data set was already reprocessed by \citet{LMD09}, who identified planet b and provided an astrometric measurement for this planet. Also, \citet{MMV10} identified planet c and a possible detection of d but without astrometric analysis. In this paper, we further the optimization of the PSF subtraction algorithm to identify the three planets b,c and d in both rolls and detail the astrometric analysis. 
In Section 2 and 3 we describe the data and the processing methods for detection, and a statistical approach to obtain astrometric estimates. 
In Section 4, we discuss orbit fitting for each planet and study the compatibility of the data with MMR scenarios that have been invoked to stabilize the system.

\section{ HR 8799 NICMOS data set from 1998}\label{sect:data}
In this paper we use the archival HST NICMOS coronagraph data set from program 7226 (E. Becklin, P.I.). This program was designed to search for massive planets or faint companions around 45 nearby young stars \citep{LBS05}. This data set was obtained with the camera NIC2, using the 0.63 arcsec diameter mask, and consists of six images in two roll orientations separated by a roll angle of 29.9 degrees (three images in each orientation) using the broad band filter F160W. \citet{LMD09} studied this data set and were able to detect and measure the astrometry of HR 8799b, using 203 reference PSFs taken from the same program to optimize the PSF subtraction using the LOCI algorithm.
Recently \citet{SSS10} achieved a complete recalibration of the NICMOS archive (through HST Cycle 15).  This recalibration provides a Legacy Archive PSF Library (LAPL) (HST AR program 11279) available as High-Level Science products integrated into the Hubble Legacy Archive\footnote{http://archive.stsci.edu/prepds/laplace/}. Improvements include the use of contemporary flats and observed darks (as opposed to epochal flats and modeled darks'), better bad-pixel correction, and ancillary information about the target (star position, magnitudes in J,H,K bands) that can help select best-matching reference PSFs from the library.  It is important to note that improvements in PSF calibration can lead to a substantial improvement in LOCI-processed data, since having random noise sources (e.g. from insufficient flat or bad-pixel calibrations) reduces the correlation between reference PSFs, and therefore the efficiency of the LOCI algorithm.
For this work, we selected 466 PSFs from the \citet{SSS10} PSF library (using contemporary flats and observed darks), which also includes all six images of HR 8799.

\section{Data processing and methods for astrometry}\label{sect:data}
\subsection{Data processing}

For each of the six HR 8799 images we first generate cubes of aligned reference PSFs by cross-correlating the diffraction spikes from the secondary mirror support structures, and by optimizing the shape of the regions used in the cross-correlation. For each HR 8799 image the corresponding reference PSF stack also includes the three images of HR 8799 in the other roll. Because the NIC2 PSF is slightly under-sampled at F160W wavelengths, we first apply a gaussian filter of two pixels full width half maximum (FWHM) before shifting the images to avoid aliasing artifacts. 

For each HR 8799 image, we apply the Locally Optimized Combination of Images (LOCI) algorithm \citep{LMD07}. 
This algorithm builds a reference PSF based on a linear combination of PSFs from a reference library, by optimizing the subtraction residuals in the least square sense. 
A remarkable feature of this algorithm is that the reference is built locally in small sections of the image, and then pieced together. This ensures a much better subtraction since speckle correlation between different PSFs depends on the speckle position within an image. The coefficients used in the linear combination are calculated in an optimization zone (``$\mathcal{O}-$zone"), and the subtraction is applied to a smaller, subtraction zone (``$\mathcal{S}-$zone"), so that the presence of a planet has minimal influence on the calculated coefficients. 
We follow the geometric implementation of the LOCI zones described in \citet{LMD07}, and we use the same geometric parameters to describe the zone shapes ($g$, $Na$), where $g$ describes the shape of the zone, $Na$ describes the number of resolution elements (speckle-equivalent areas) in the $\mathcal{O}-$zone. A third parameter $dr$ is used by \citet{LMD07} to describe the width of the $\mathcal{S}-$zone. Even when exploring a vast ($g$, $Na$, and $dr$) parameter space, we were not able to detect all three planets in every image. As a consequence we implemented a series of improvements to the initial algorithm:
\begin{itemize}
\item{\it Zone geometry}: we explored several variations of LOCI zone geometries, including the classical LOCI geometry \citep{LMD07}, large annulus zones \citep{LMD09}, small $\mathcal{S}-$zone of one resolution element \citep{MMV10}. Our best results are obtained using a $\mathcal{S}-$ zone composed of a single pixel. We therefore discard the third parameter $dr$, as our $\mathcal{S}-$zone always consists of a single pixel.
\item {\it Masking}: the single-pixel $\mathcal{S}-$zone is excluded from the $\mathcal{O}-$zone to preserve better the signal from the planet PSF. 
In addition, we also mask out a few pixels from the $\mathcal{O}-$zone surrounding the $\mathcal{S}-$zone. With a masking zone comparable to a planet PSF size this insures that the signal from a possible planet would not impact the correlation matrix and therefore this preserves throughput \citep{MMV10}.
\item {\it $\mathcal{S} -$ zone centering}: in the original LOCI, the $\mathcal{S}-$zone is situated at the inner-edge of the $\mathcal{O}-$zone (see Figure 1 in \citet{LMD07}). We added a parameter to shift the radial position of the $\mathcal{S}-$zone within the $\mathcal{O}-$zone so that the pixels surrounding both sides of the $\mathcal{S}-$zone are included in the optimization region.
\item {\it Selection of optimal reference images used for each $\mathcal{O}-$zone}: LOCI is more efficient when the number of constraints (e.g. the number of PSF cores, $Na$, in the optimization zone) is equal to the number of degrees of freedom (e.g. number of references that are linearly independent in the $\mathcal{O}-$ zone), \citep{MMV10,Pueyo11}. For each $\mathcal{O}-$zone we introduce a parameter that optimizes the number of references images as a function of their degree of correlation. If for example there is a background source in one of the reference PSFs, this reference will be automatically rejected from the reference stack for any O-zone that includes the stellar companion because of the poor correlation with the target.  Note that this same PSF might still be usable (and included in the reference stack) for O-zones that do not encompass the stellar companion.

\item {\it Conditioning of the correlation matrix}:  First, prior to the LOCI-reduction, we apply a classical PSF subtraction both to the target image and reference stack,  i.e. we take a median through the aligned reference cube, and then subtract this median image with the target and with each reference image. Thus, we are able to subtract the truly static modes that are significant contributors to the poor conditioning of the correlation matrix. In some cases, the correlation matrix is still poorly conditioned, in which case we use matrix regularization.
\end{itemize}

In our study of the LOCI algorithm, we note that the most important parameter is the zone geometry. This affects both self-subtraction and signal to noise ratio (SNR). For example, very large zones correspond to little planet self-subtraction because the impact of the planet in the correlation matrix is diluted in the large optimization region.  Masking is also very important, as this greatly reduces planet self subtraction since the majority of the planet PSF is excluded from the correlation matrix. However, if the mask is large compared to the PSF size, it also affects the SNR because the speckles close to the planets are not included in the correlation matrix, and therefore the algorithm does not subtract the PSF well in the immediate vicinity of the planet. We find matrix conditioning becomes important when we use a large number of references for a given O-zone. The regularization of the ill-conditioned correlation matrix can greatly affect the quality of the subtraction result. Other parameters such as S-zone centering have a lesser impact on the results.

\begin{figure*}[htbp]
\center
\resizebox{0.8\hsize}{!}{\includegraphics{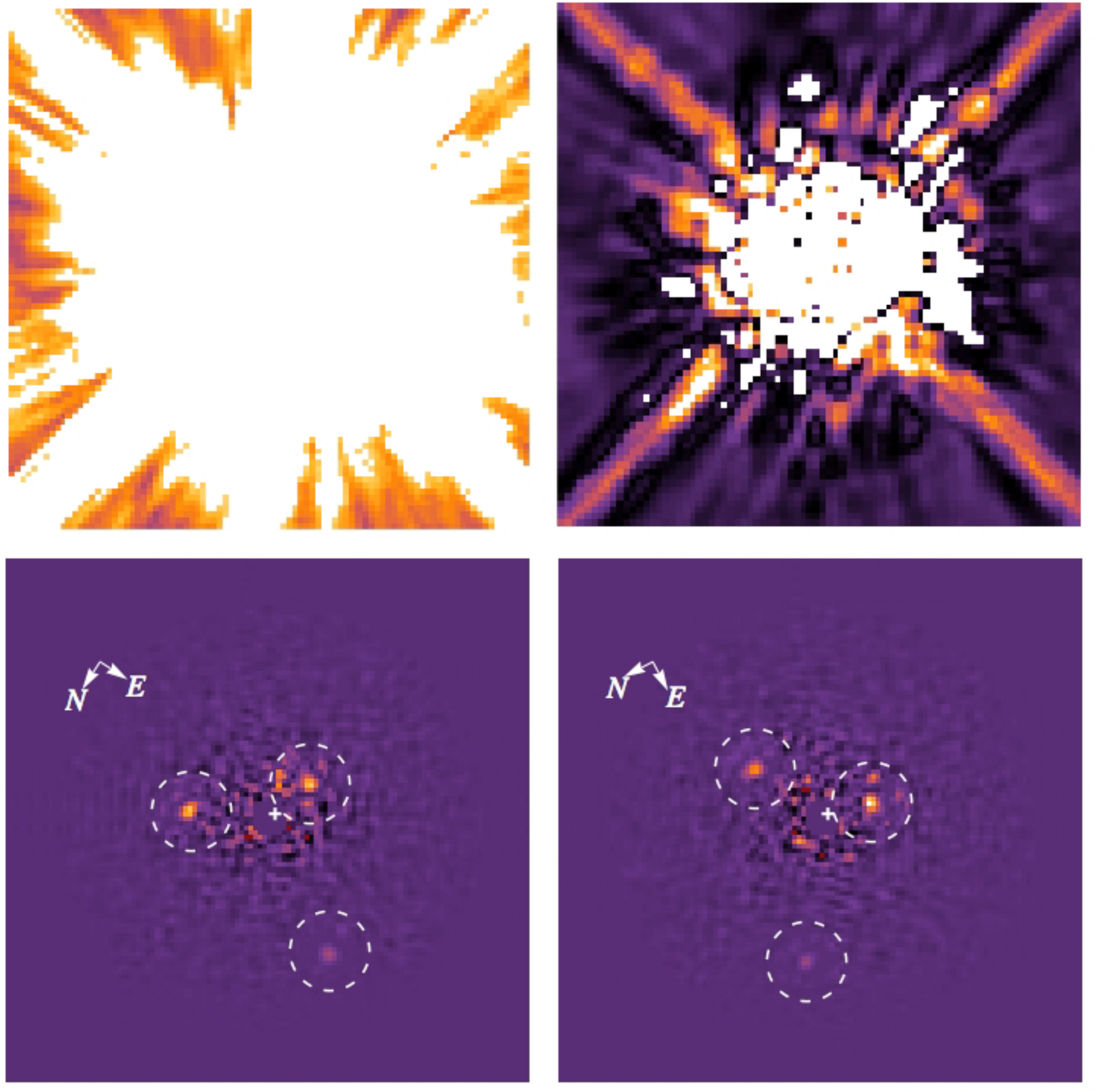}}
 \caption{Initial coronagraphic image of HR 8799 (top left), Classical PSF subtraction (top right) and final processed images showing the three planets HR 8799b, c and d in both rolls in the HST NICMOS data from 1998. The figure illustrates the process followed in this work: we first used LOCI over the entire image to identify potential candidates for all three planets at lower detection levels. We then concentrated the work on small regions around each planet (indicated by the white circles) to improve detection and astrometry, using a pixel-wise LOCI reduction with an exclusion region optimized to avoid astrometric biases. The figure shows a composite image with the preliminary LOCI in the background and with the more refined optimization shown inside the white circles. We used the median of the best images in these small regions to improve signal to noise in this image. See also Figure \ref{Fig:stamps} for individual examples of the data. The improvement over the classical PSF subtraction is an order of magnitude in contrast. All four images are displayed on the same scale, hence the apparent saturation of the initial coronagraphic image}\label{fig1}
\end{figure*}

Our approach involved several successive LOCI reductions. First, we used a classical LOCI implementation to identify b and c in some of the images, and a low-detection candidate for d in one roll. Then we refined LOCI optimizations in small regions of interest around each planets using a single-pixel $\mathcal{S}-$zone LOCI with an exclusion zone surrounding the $\mathcal{S}-$zone. We then successfully detected all three planets in both rolls. 
We illustrate this process and show a final composite image in Figure \ref{fig1} where the background corresponds to one of the initial LOCI reductions, and where the regions inside the white circles correspond to the more refined LOCI reductions around each planet position. 
In this paper we develop methods to characterize the astrometry and prove that all of our detections are truly companions.

We also searched for planet e in the predicted region based on the 2010 discovery data \citep{MZK10}, and identified a possible candidate, but its position is not consistent in both orientations so it is likely to be a residual speckle. Moreover, HR 8799e would be just outside the edge of the coronagraphic mask where the NICMOS PSF has known glints \citep{STS98}, which are due to the imperfections in the coronagraphic hole. While these glints are in principle suppressed by LOCI, they add potential false alarms in this region.

Compared to the results by \citet{LMD09} with the same data set, our more detailed LOCI optimization brings improvement in contrast at short separations for planets c and d, and we obtain a comparable detection for b. 
In their images planet c is visible in one roll, and although d is not detected, there are bright pixels at the right location. We therefore estimate the improvement in SNR by a factor of $\sim2$ in single LOCI reduced images since our single images of d have an SNR of the order of 4, to be compared with the image by \citet{LMD09} using a single LOCI-reduction. Because our method also includes the exploration of a very large LOCI parameter space, detectability is further improved by combining the images as shown in Figure 1. 

\subsection{Astrometric characterization}

The determination of the orbital motion requires relative astrometry measurements between the star and the planets. Given the NIC2 pixel size of 76 mas, sub-pixel astrometric accuracy is needed to provide significant constraints on the orbits. Such measurements are extremely difficult to make in the case of the HR 8799 planets, since these objects are much fainter than speckles in the coronagraphic PSF. An aggressive PSF subtraction is therefore required, which can potentially bias the astrometry. Moreover, the relatively low SNR obtained even after subtraction (in particular for our detection of d) requires careful estimation of error bars and potential biases in order to fit the orbits properly.
In the following sections, we detail each steps we followed for the astrometric characterization of the HR 8799 planets. These steps include measuring the star astrometry, generating LOCI-reduced images, studying the astrometric statistics as a function of signal to noise ratio (SNR), and in particular the agreement between both rolls, estimating our astrometric biases and residual errors, and testing the possibility of false alarms.

\subsubsection{Star position}

Because the star is occulted by the coronagraph,  and since there is not a differential measurement that would eliminate any systematic biases, its position has to be calculated independently from the planets. This problem has been solved for the next generation of instruments, by creating satellite diffraction spots \citep{SO06,MLM06}. For the star position measurement, we follow the method described by \citet{LMD09} using a Tiny Tim Model of the NICMOS PSF \citep{HK97} to cross-correlate the diffraction spikes. Assuming a coordinate system where zero is located a the bottom left corner of the image we find (72.46, 211.46) and (72.46, 211.47) for each roll, which is within 0.1 pixel of the estimates by \citet{SSS10} (72.41, 211.56), and also in excellent agreement with the measurement used by \citet{LMD09} (72.50, 211.42) (Private Communication). We therefore estimate the error on the star position to have a standard deviation $\sigma=0.1$ pixel.

\subsubsection{Post processing astrometric scatter}

\begin{figure}[htbp]
\center
\resizebox{0.6\hsize}{!}{\includegraphics{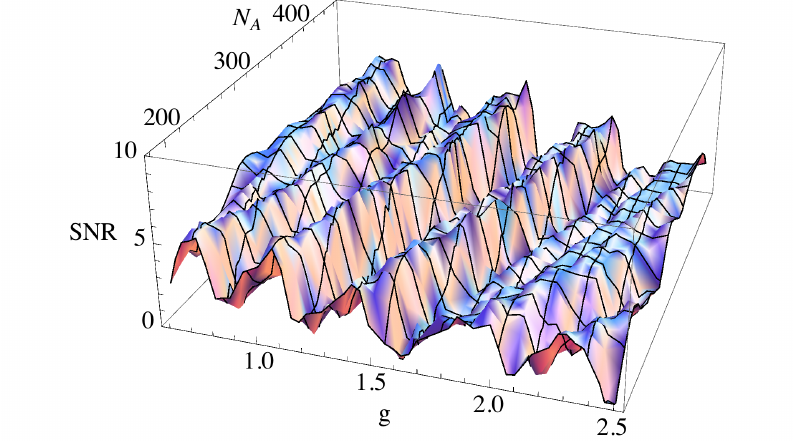}}
 \caption{SNR of planet b as a function of LOCI parameters $g$ and $Na$ that define the shape of the optimization regions. Note that there are several ripples where the SNR is high. Instead of working with a single high-SNR image and obtain the astrometry from it, we use the SNR as a criterion to select the best images and obtain astrometry from many measurements on a large number of LOCI-reduced images. Note that there are also other parameters used in the LOCI optimizations that also lead to SNR variations (position of the S-zone within the $\mathcal{O}$-zone, number of PSF references used).}\label{Fig:SNRripple}
\end{figure}

We first explored a vast parameter space for LOCI to find the algorithmic parameters yielding the best detectability of each planet. In Figure \ref{Fig:SNRripple} we show the evolution of the signal to noise ratio (SNR) for planet b as a function of the two LOCI parameters $g$ and $Na$ \citep{LMD07}, which control the geometry of the optimization zone. The SNR does not exhibit a well-identified maximum for a unique set of LOCI parameters. 
As a consequence we used a grid of LOCI parameters and generated 2,100 LOCI-reduced images for each of the six HR 8799 images (total of 12,600 images) using a single-pixel $\mathcal{S}$-zone LOCI with a $3\times3$ masking box centered on the $\mathcal{S}$-zone, identified below as the least biased algorithm among the variations we tested (see below). Our algorithm parameterization includes the two LOCI parameters $g$ and $Na$ \citep{LMD07} that determine the LOCI zone geometry, a parameter to select the number of PSFs to include in the reference for each $\mathcal{O}$-zone, and a parameter for the radial position of the $\mathcal{S}$-zone relative to the $\mathcal{O}$-zone. 

For each reduced image, we used matched filtering to measure the signal and the astrometry. This approach is optimal in LOCI images where speckle noise has been sufficiently suppressed to leave residual gaussian noise. This was verified by \citet{LMD09} who found that the residual noise in the optimization region around the b planet closely follows a Gaussian distribution. We cross-calibrated various match filter templates (from real data, Tiny Tim or direct simulation) with a perfect PSF of known position, to avoid any biases from this measurement. Our best results are obtained with an adhoc PSF obtained with a simulation for the matched filter, verified through the fake planet injection astrometric analysis (described below in Section 3.2.4) to give the most accurate position of the planet PSF post LOCI reduction. 
Because the astrometric position is not perfectly known, and because some sets of LOCI parameters produce low-contrast images, it is possible that the matched filter settles on a speckle close to the true planet. We eliminate these images if the match filter does not detect anything within a three-pixel diameter area around the planet position estimate, considering that this is a very conservative estimation based on direct visual analysis of the planet candidates. For any other images where the matched filter detects something within this three-pixel diameter area, we consider a positive detection and record its SNR and astrometry.
 Here we are only interested in astrometric characterization and not detection, so we define a local SNR, where ``local'' means that we are only optimizing the PSF subtraction in a small region centered around the planet (see the white circles in Figure 1).  We therefore designed our SNR metric to only consider the noise in a small annulus centered on the planet (the outer radius of this annulus is given approximately by the white circles in Figure 1).

\begin{figure*}[htbp]
\center
\resizebox{\hsize}{!}{\includegraphics{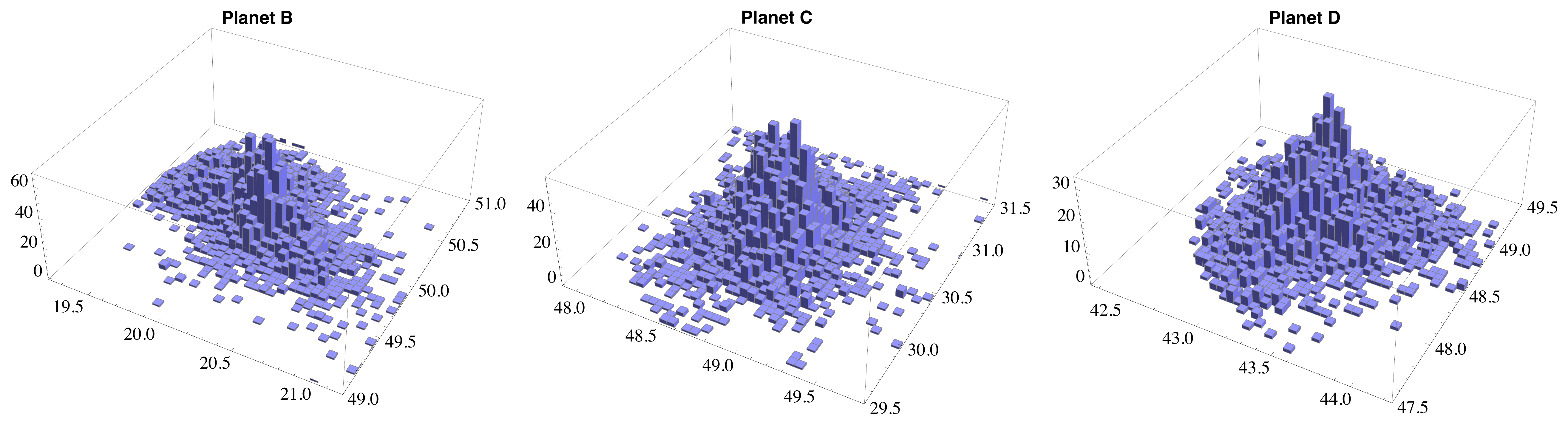}}
 \caption{3-D histogram of the astrometric distribution for all detections of each planet in the best roll exploring a large number of sets of LOCI parameters (2,100 LOCI-reduced images per HR 8799 image).
 The X and Y axes correspond to the x and y measured positions of the planet in pixels in a $2\times2$ pixel region, where the origin corresponds to the actual measured position of the real planet. Results are included in the statistics if the matched filter detects a source within an area of three-pixel diameter centered on the planet position estimate. These figures include all measurements made in the data without any selection criterion based on SNR. Note the dispersion of the astrometric data. By exploring a large number of parameters for LOCI we randomize some of the astrometric errors due to the algorithm itself and speckle noise and mitigate the astrometric error we would obtain from a single LOCI-reduced image. 
 }\label{Fig:3Dhistogram}
\end{figure*}

As the LOCI parameters vary we observe a scatter for the measured astrometric positions. This scatter is due to a combination of two effects: the aggressive PSF subtraction can alter the shape of the planet PSFs, and the residual speckles in the LOCI-reduced images, which depend in part on the algorithm parameters, impact the match filter measurement. 
This astrometric scatter is illustrated in Figure \ref{Fig:3Dhistogram}, which shows the histogram of the X and Y astrometric positions for all the positive ``detections'' in the sense defined above. 
The scatter shown in Figure \ref{Fig:3Dhistogram} spans over a $2\times2$ pixel region, but some of these measurements correspond to very low-SNR values. Therefore it is not possible to use this result directly to obtain an astrometric position and error bar for each planet. Moreover, there is no guarantee that the scatter is not biased because of the effect of the LOCI algorithm on the planet PSF shape, and of residual static speckles on the match filter measurements. 

\subsubsection{Reducing the astrometric scatter to sub-pixel level}

\begin{figure*}[htbp]
\center
\resizebox{\hsize}{!}{\includegraphics{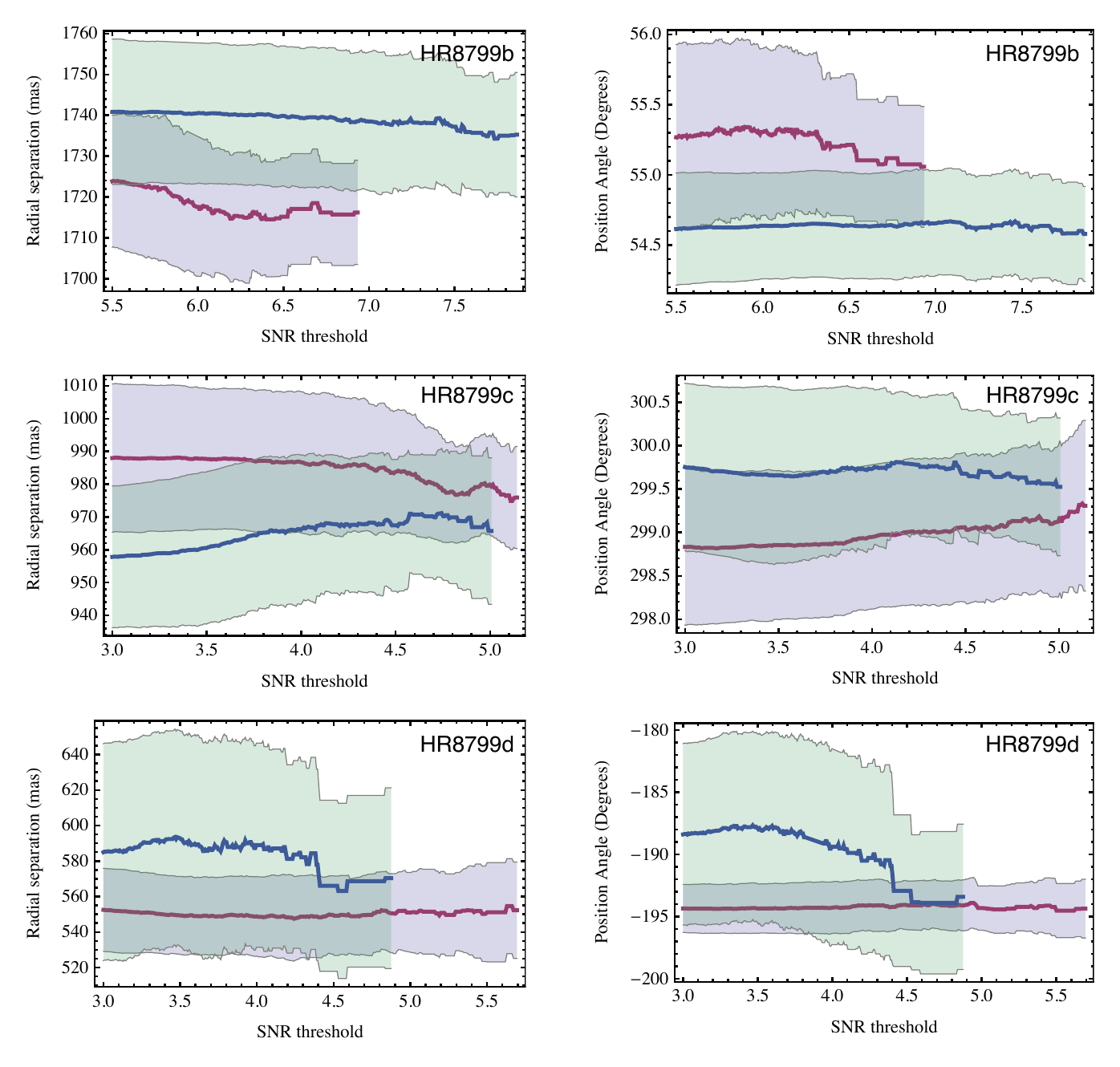}}
 \caption{For each planet b, c and d, we show the mean (thick lines) and standard deviation (shading on both side) of the astrometric measurements in each roll (  roll 1 in blue with green shading, roll 2 in red with gray shading) for LOCI subtracted images with SNR above a minimum SNR threshold (shown on the x-axis). The standard deviation is obtained from the astrometric measurement of the LOCI-reduced images of the six HR 8799 images (three in each roll). 
 As the minimum SNR increases, fewer images are included in the statistics. The maximum SNR threshold displayed in these plots corresponds to the SNR for which there are 10 images with higher SNR than the threshold. 
As the SNR increases, the astrometric agreement tends to improve from roll to roll. We use a minimum SNR of 6.5 for planet b, and 5.0 for planet c and 4.5 for planet d to calculate the final astrometry by combining the measurements in both rolls for these SNR threshold values. The discrepancies between both rolls are indicative of the residual astrometric biases (speckle noise, star position, systematics) that cannot be calibrated. These effects are accounted for by measuring the standard deviation of measurements made in both rolls. Note that we use a local SNR where the noise is calculated in a small region around the planet since we are only interested in astrometric characterization and not in detection.
 }\label{Fig:SNRthreshold}
\end{figure*}

\begin{figure}[htbp]
\center
\resizebox{0.6\hsize}{!}{\includegraphics{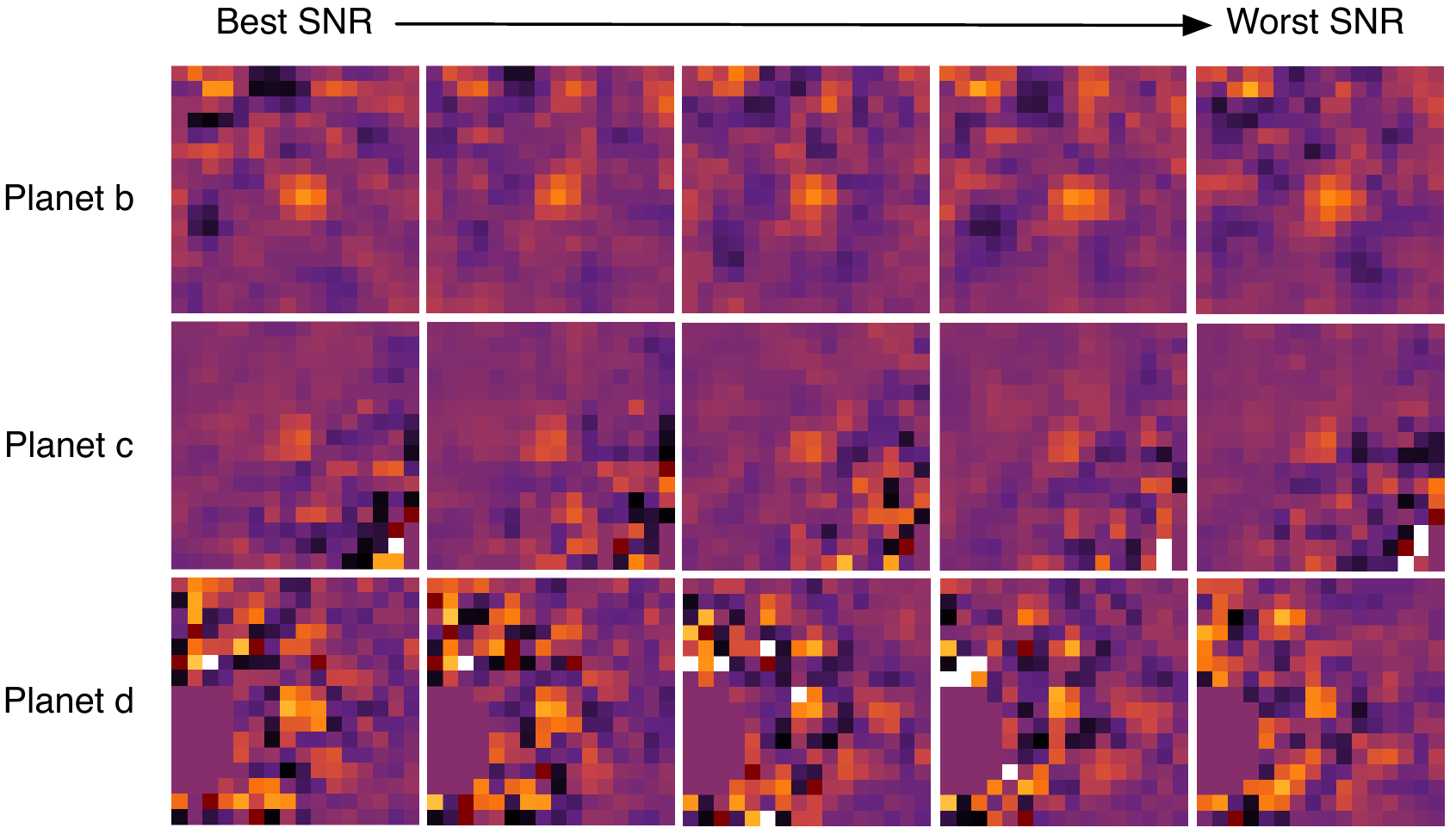}}
 \caption{The figure shows a few examples of LOCI reduced data that are used in the statistics for the astrometry in one roll. For each three planet b, c, and d (each row) we show five images corresponding to different sets of LOCI parameters. These images are ordered by decreasing SNR from left to right. The left images correspond to the single image with highest measured SNR, and the right ones to the images with lowest SNR used in the statistics. These five images span evenly the range of 293 images for b (2\% of the best images in roll 1), 23 images for c (0.3\% of the best images in roll 2) and 66 images for d (0.6\% of the best images in roll 2). Our astrometric measurements are obtained using matched filtering on all of these images. We claim a minimum SNR of 6.5 for planet b, 5.0 for planet c, and 4.5 for planet d. In this paper, we are concerned mostly with the astrometry of these planets, and thus the SNR is measured in a region very local to the planet position. If we were to measure the SNR globally for each planet, we would have lower values of SNR for each planet. 
 The inserts in Figure 1 inside the white circles correspond to a median of these series of best images. Also, note how varying LOCI parameters help randomize the shape of the planet and the speckle noise properties around it. }\label{Fig:stamps}
\end{figure}

We reduce astrometric scatter by using the SNR derived from our local matched filter as a proxy to study the statistical relevance of the astrometric estimates in a given image. For each planet, we select an ensemble of LOCI-reduced images that have an SNR above a minimum SNR threshold value, and we compute the mean and standard deviation of the astrometric position in this ensemble. Figure \ref{Fig:SNRthreshold} shows how both quantities evolve as a function of the SNR threshold for both rolls. The measured astrometric positions of the three planets overlap between both rolls within their error bars. Also, the mean of the astrometric measurement in each roll tends to converge for high SNR values, and their scatter is reduced. While it is tempting to derive astrometric estimates and error bars from the ensemble with the highest SNR, there is a trade-off between the lack of statistics for the highest SNR values, and using lower SNR images in the statistics. We thus select images with minimum SNR of 6.5 for b ( 2\% best images), 5.0 for c ( 0.3\% best images), and 4.5 for d ( 0.6\% best images) to obtain the mean and standard deviation of the astrometric positions, combining measurements from both rolls. Note that according to Figure \ref{Fig:SNRthreshold}, the astrometric result does not depend strongly on the choice of the SNR threshold. In Figure \ref{Fig:stamps} we show individual examples of the LOCI-reduced images for each planet with SNR ranging evenly from the the minimum SNR threshold to the maximum SNR for the series.

\subsubsection{Estimating astrometric positions and potential biases}

The value of the mean astrometric position derived using the technique above can be potentially biased, and we study the astrometric errors by carrying out the same method to a reference PSF with fake planets injected at the same location and brightness as HR 8799. There are two possible sources of astrometric errors and biases. 

The first source is associated with the LOCI implementation. If the algorithm is too aggressive (e.g. without using masking) LOCI subtracts part of the planet signal, therefore altering the planet PSF shape and biasing the measurement of its position. Using a classical LOCI implementation without the masking technique, we measured astrometric biases up to about 2/3 of a pixel ($\simeq 50$ mas). With the LOCI masking technique introduced by \citet{MMV10}, each subtraction region is excluded from the corresponding optimization zone so that the $\mathcal{S}-$zone pixels do not impact the correlation matrix.
Therefore this technique mitigates the astrometric biases caused by a partial subtraction of the planet PSF. 

The second source of astrometric error is unavoidable and is due to residual speckles very close to the planet position, which can bias the matched filter measurement (or any other astrometric method e.g. gaussian fitting). Residual speckles depend upon the number of references images used, how well they are correlated with the target image, and on the choice of LOCI parameters. Large exclusion zones limit the LOCI performance close to the planet by excluding the corresponding pixels from the correlation matrix. This in turn yields increased speckle noise around the planet position, and associated astrometric errors. Varying the LOCI parameters introduces some speckle diversity that randomizes the astrometric errors at some level, but there are speckles that remain static even with different LOCI parameters. With large exclusion zones (several resolution elements) we measured biases of the order of 1/2 pixel.

There is therefore a trade-off between the aggressiveness of the LOCI subtraction, and the amount of residual speckle noise close to the planet.
After testing many LOCI implementations with various zone geometries (with and without masking) and exclusion zone sizes, we selected a LOCI implementation that produces low overall astrometric bias for this data set, while enabling excellent PSF subtraction. The NICMOS PSF at F160W wavelengths has a FWHM of $\simeq2$ pixels, thus we selected a single-pixel $\mathcal{S}-$zone LOCI with a $3\times3$ exclusion region around each $\mathcal{S}-$zone. 
This ensures that most of the core of a potential planet PSF is not included in the correlation matrix (low astrometric bias), while speckles very close to the planet core are (aggressive subtraction, and lower residual speckle noise). 
This result may vary for other datasets, in particular for different detector samplings. 
In addition, this LOCI implementation was the only one we found that was capable of detecting all three planets successfully in both rolls with roll to roll agreement consistent with the error bars.

We performed our quantitative study of astrometric errors and biases by injecting fake planets in two PSFs from the LAPL archive \citep{SSS10} with very similar stellar properties and same exposure time as HR 8799. For the fake planets brightness we used photometric values based on their H-band and F160W photometry from \citep{MMB08,LMD09}. Our fake planets are slightly fainter ($\sim\times$1.5) than the real ones, which makes our estimations more conservative. 

\begin{figure}[htbp]
\center
\resizebox{0.35\hsize}{!}{\includegraphics{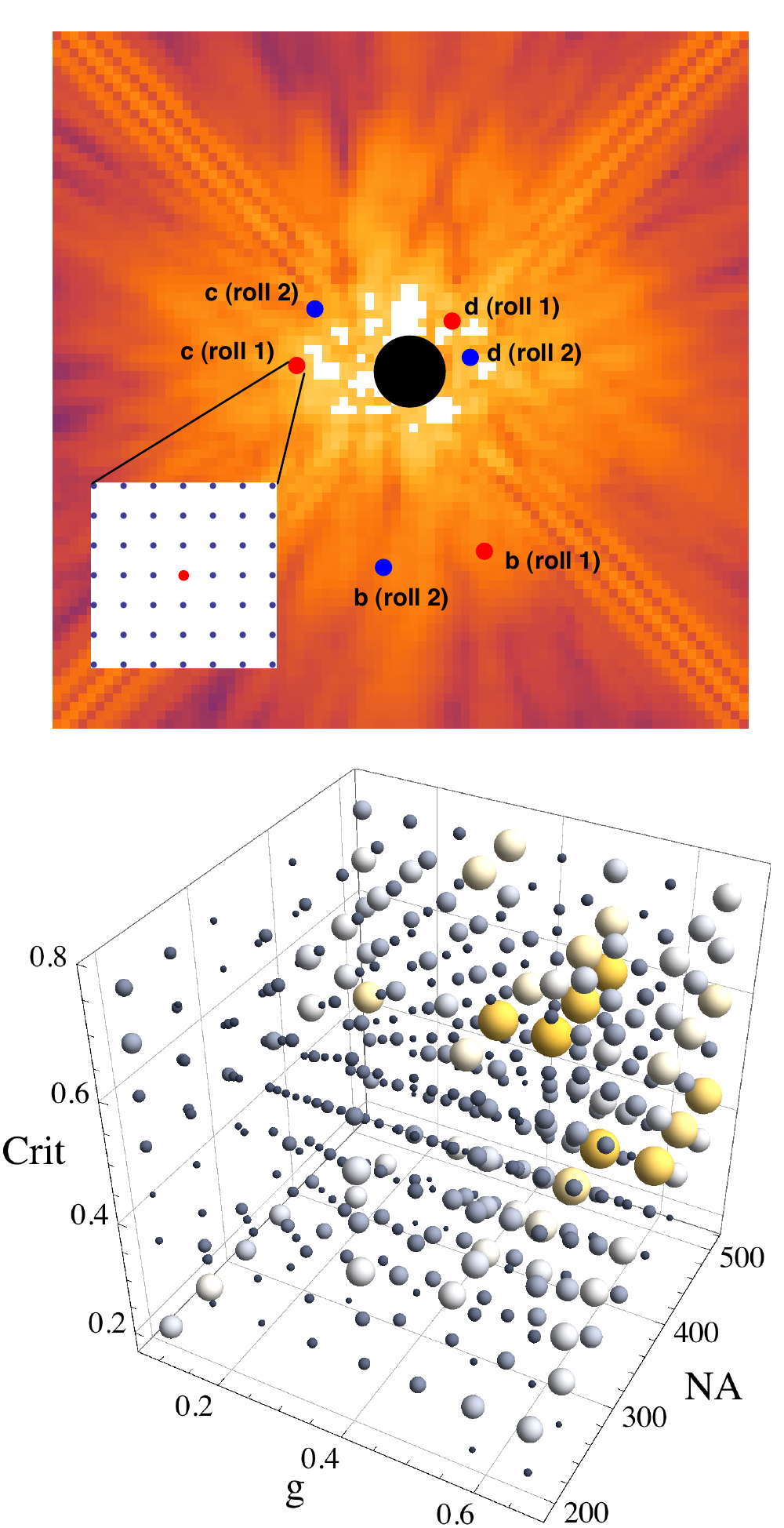}}
 \caption{Study of astrometric errors as a function of LOCI parameters. We generate fake planetary systems by injecting fake planets on a $7\times7$ grid, centered on the measured position of the real planets (left). Note the position of planet c in the roll 2 position and planet d in the roll 1 position, located on the diffraction spikes of the star PSF (the locations where we measure the largest bias; see Table \ref{tableBiases}). For each position on the $7\times7$ grid, we explore a LOCI parameter space of $\sim500$ parameters, thus generating on the order of 100,000 LOCI reduced images for this study, $\sim50,000$ for each star PSF (2 rolls). For each planet and each LOCI parameter, we measure the error between the true position and the measured position. The 4-D plot (right) studies the astrometric errors for each LOCI parameter tested. Each bubble corresponds to a set of LOCI parameters (only 3 parameters of our algorithm are displayed here: the $g$ and $Na$ geometric parameters from \citet{LMD07}, and a criterion (labelled ``crit'' in the figure) used to select the number of reference frames to be included in each optimization zone). The fourth dimension (bubble diameter), is proportional to the number of fake planet positions (between 1 and 49) for which the astrometric error is smaller than 0.1 pixels for planet b in the roll-2 position. Note that the top right corner of this parameter space typically leads to better astrometry for this radial seperation and position angle in this PSF. Although the general behavior is consistent with another PSF at the same radial separation and position angle, there is no ideal value of the parameter space that guarantees the best astrometry across all positions of the PSF, and therefore we use the statistics to define the planet astrometric measurements instead of relying on a single image. Furthermore, when comparing the two PSFs tested for astrometric biases, we find that not all positions from PSF to PSF have consistent astrometric biases for similar parameters.
 }\label{Fig:FakePlanetInjection}
\end{figure}

Our choice of using two PSFs is designed to test whether some of these astrometric errors are consistent for different PSFs and thus whether they can be calibrated.  We verified that there is no ideal set of LOCI parameters that provides unbiased astrometry at the sub-pixel level for all three planet positions. 
For each PSF and each astrometric position (b, c and d), we injected fake planets on a uniform $7\times7$ grid spanning the size of a full pixel centered on each measured planet position.
We therefore generated 49 fake planetary systems in two different orientations to simulate each roll (total of 98 fake planetary systems). 
We then applied the same methodology as for the real HR 8799 data, albeit using a sparser parameter space with $\sim500$ LOCI parameters for each fake planetary system (as opposed to 2,100), generating of the order of 100,000 LOCI reduced images, and then studying the astrometric errors and measurement biases for each fake planet position. This process is illustrated in Figure \ref{Fig:FakePlanetInjection}. 

In our study, we find that the astrometric biases vary with the position of the planet in the star PSF. This is most apparent when the planet PSF is located on the diffraction spikes of the star PSF. Typically we find low biases on planet positions not located on the diffraction spikes, and higher bias for those that land on the diffraction spikes (Table \ref{tableBiases}). Furthermore, we find the biases are not sufficiently consistent from PSF to PSF, and thus these results cannot be used to calibrate the actual HR 8799 reduced images. 
Planet d in the roll 1 position and planet c in the roll 2 position are located on the diffraction spikes of the star PSF (Figure \ref{Fig:FakePlanetInjection}), which produces astrometric biases on the order of 25-40 mas. For all other positions we find that we have typical astrometric biases on the order of $0-15$ mas ($\lesssim 0.2 $ pixel), which remain small compared to the error bars (18 mas for b, 22 mas for c, and 29 mas for d, see Table \ref{tableBiases} for complete quantitative analysis). In these final astrometric  results we do not include the statistics from planet d in the roll 1 position and planet c in the roll 2 position, to avoid the biases observed in simulations due to the presence of the diffraction spikes. Nevertheless, in the real HR 8799 data, we find a good agreement between both rolls for the mean astrometric measurements for all three planets (see Figure \ref{Fig:SNRthreshold}), an indication of the likely low bias of our astrometric measurements. 

\begin{figure*}[htbp]
\center
\resizebox{\hsize}{!}{\includegraphics{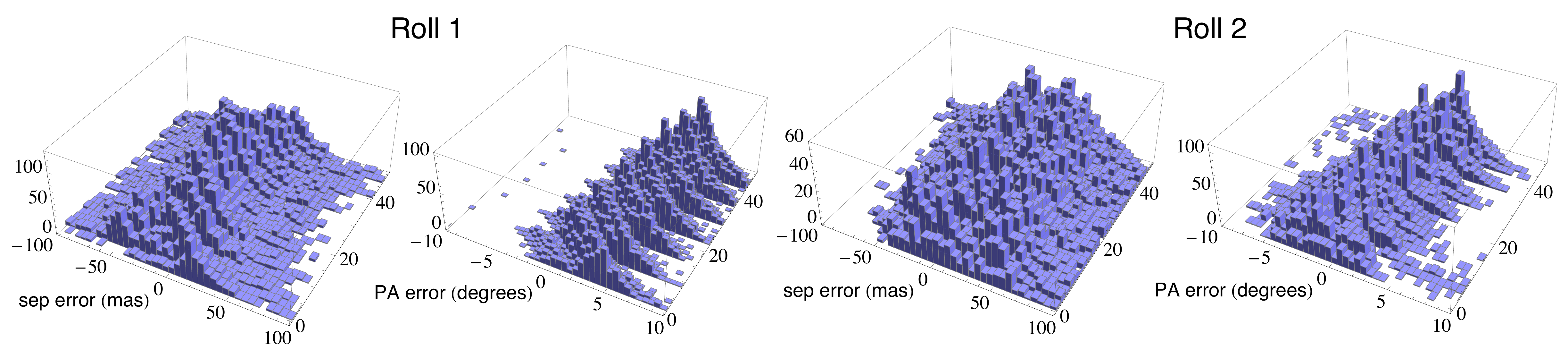}}
 \caption{Planet D Radial Seperation and Position Angle for Both Roll Orientations. See Table \ref{tableBiases} for the measurement of the biases from these statistics. For each roll orientation we show two 3D plots. Each plot contains 49 histograms of the astrometric errors for the radial seperation/position angle of d-planet. The X-axis gives the measure of the error from the true position. The Y-axis, with range 1-49, corresponds to one of the 49 tested positions for astrometric bias in the $7\times7$ grid, spanning the area of one pixel (see Figure \ref{Fig:FakePlanetInjection}). Note the larger bias for the position angle of planet d in Roll 1, due to the location of the planet PSF on the diffraction spike of the star PSF. We note for the other PSF tested, the bias for planet d at the roll 1 position is large in the ortho-radial direction (\ref{tableBiases}), thus these biases are not consistent from PSF to PSF and cannot be calibrated. We do not include these positions in the final astrometry. Other positions show low bias (0-15 mas) compared to the error bars. }\label{Fig:3Dhisto}
\end{figure*}

We provide an example of the astrometry error statistics performed on one PSF for the separation and position angle of planet d in roll 1 and roll 2 in Figure \ref{Fig:3Dhisto}. Each histogram in the 3D plot corresponds to the error on the separation and position angle for each of the 49 fake planets injected on the $7\times7$ sub-pixel grid. The position angle biases are small in both rolls (-0.15 and 0.17 pixels), but there is a large bias on the angular position in Roll 1 (0.47 pixel) compared to the low bias in roll 2 (0.04 pixel). We indentified this large bias in Roll 1 to come from the location of planet d on the diffraction spike of the star PSF (see Figure \ref{Fig:FakePlanetInjection}). 

Since the potential astrometric biases cannot be calibrated, and we find we have very low biases outside of the diffraction spiders of the star PSF, our strategy is to use only the measurements obtained where the position of the planet PSF does not fall on the diffraction spikes of the star PSF, even though we have excellent astrometric agreement from roll to roll for all planets. Thus, for the final astrometric numbers, we use the measurements for planet b in both orientations, planet c in roll 1, and planet d in roll 2. To be conservative, we define the astrometric error bar by combining the overall scatter of astrometric measurements made from both roll orientations, and use the largest error between the separation and position angle, and force the error of the position angle and separation to be equal to the larger of the two measurements. Ideally, the best approach to minimize astrometric biases would be to use a large number of roll orientations in order to average possible biases over several positions (e.g. as it can be done using ADI on the ground). 
The uncertainty on the absolute north orientation for NICMOS is of the order of 0.1 deg. We corrected for the rectangular geometry of the pixels when projected on the sky, and higher distortion terms are of the order of 1 mas and therefore not significant compared to the other errors.  Also, in order to account for uncalibrated systematic errors on the absolute North orientation between telescopes, we add a one-sigma error of 0.5 degree to our data for the orbit fitting (shown in a separate column in Table \ref{tableResults}).

\begin{table*}
\begin{center}
\caption{Measured astrometric biases based on fake planets simulations\label{tableBiases}}
\begin{tabular}{ccccc}
\tableline\tableline 
\tableline
 Planet  & Roll 1 sep (mas) & Roll 1 az (mas) & Roll 2 sep (mas) & Roll 2 az (mas) \\
\tableline
 HR 8799b & -13 $\|$ -10 & 7 $\|$ 9 & -7 $\|$ 3 & 1 $\|$ 5 \\
 HR 8799c & -5 $\|$ -16 & 11 $\|$ 10 & \textbf{14 $\|$ 25} & \textbf{16 $\|$ 4}\\
 HR 8799d & \textbf{-12 $\|$ -43} & \textbf{36 $\|$ 4} & 13 $\|$ -16& 3 $\|$ 10\\
\tableline
\end{tabular}
\tablecomments{This tables shows the measured astrometric biases using fake planets injected in two different PSFs chosen from the reference library that match the HR 8799 PSFs closely. These results are based on $\sim$100,000 LOCI simulated images. The injected planets were $\sim1.5$ times fainter than the real planets, which reduces SNR and leads to a very conservative estimation of the biases. 
For each roll, the table gives the bias for each PSF. The biases are not fully consistent between PSFs, and therefore we do not attempt calibration.
The large biases (identified in bold) correspond to positions of planet c and d on top of the diffraction spikes (\ref{Fig:FakePlanetInjection}). Because of the potential biases at these locations we do not include the corresponding measurements in the final astrometry.
}
\end{center}
\end{table*}

\begin{table*}
\begin{center}
\caption{Astrometric measurements.\label{tableResults}}
\begin{tabular}{lcrrrrr}
\tableline\tableline
Reference & Planet & sep (mas) & PA (deg) & $\sigma$(mas) & SNR & fraction\\
\tableline
Lafreni\`ere et al. 2009 & HR 8799b & $1721 \pm 12$ & $55.1\pm 0.4$ (13mas)&n/a & n/a & n/a\\
This work & HR 8799b & $1738 \pm 18$ & $54.7\pm 0.4$ (13mas) & 23 & 6.5 & 2\%\\
This work & HR 8799c & $966 \pm 22$ & $300\pm 0.8 $ (13mas) & 24 &5.0 & 0.3\%\\
This work & HR 8799d & $549 \pm 28$ & $166\pm3.0 $ (29mas) & 30 & 4.5 & 0.6\%\\
\tableline
\end{tabular}
\tablecomments{Astrometric measurements (separation and position angle from North) for all three planets with comparison with Lafreni\`ere et al. (2009). 
Measurements for b includes both rolls, while only Roll 1 for c and only Roll 2 for d because of the presence of the diffraction spikes. 
The column $\sigma$ gives the overall error used in the orbit fitting, adding a 0.5 degree, one-$\sigma$ error to account for absolute North calibration difference and other systematics between observatories.
The SNR value corresponds to the minimum SNR threshold to include images in the statistics, and the last column provides the corresponding fraction of the total images that are included in the statistics with SNR greater than the SNR threshold.}
\end{center}
\end{table*}

\subsubsection{Studying false alarm probability for planet d}

Although the detection of planets b and c have a high significance, the detection of planet d is of low-significance in individual LOCI-reduced images (Figure \ref{Fig:stamps}).
However d has the shortest period and has the most orbital motion, thus it is most interesting scientifically. We therefore studied in detail the case of d to eliminate the possibility of a false alarm, and used three independent tests: astrometric analysis of bright speckles in the immediate vicinity of d, estimation of the probability of false alarm, and photometry.

\begin{figure}[htbp]
\center
\resizebox{0.3\hsize}{!}{\includegraphics{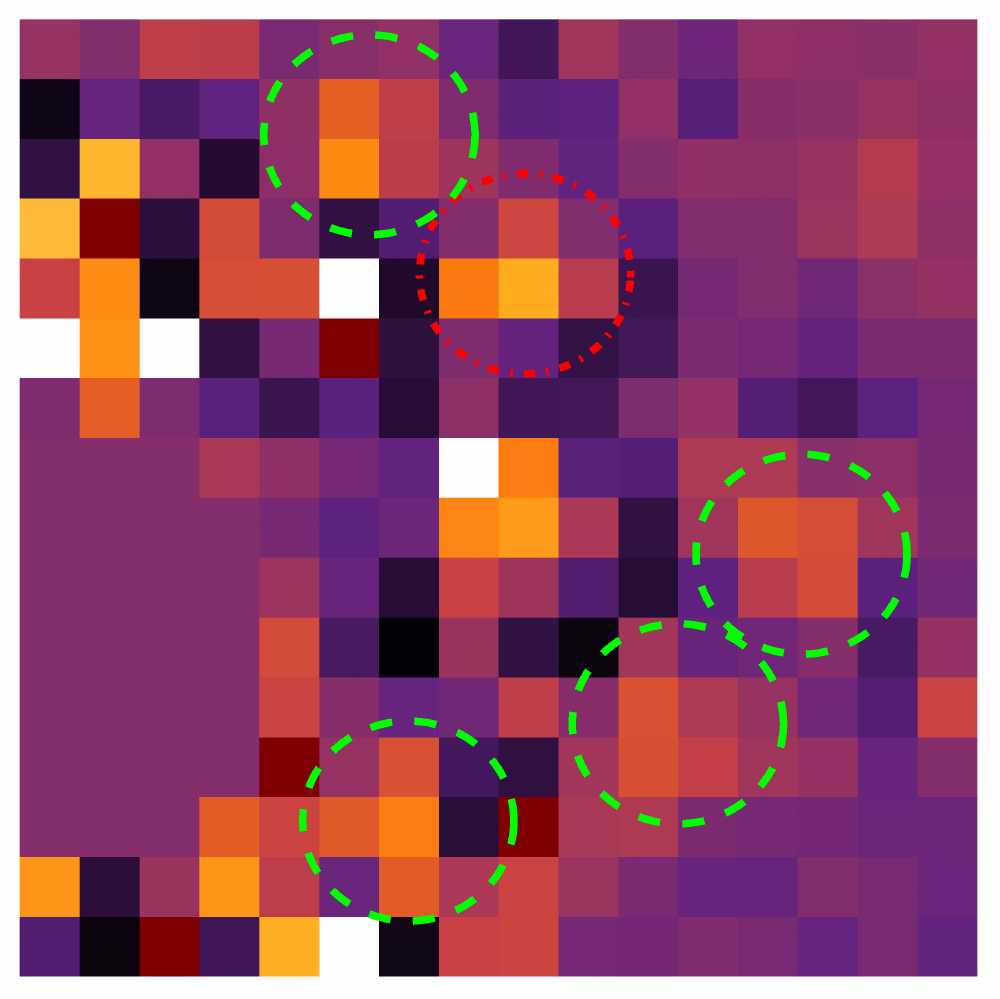}}
 \caption{Example of a LOCI-reduced image showing planet d at the center, and some residual speckles. We select the five most prominent speckles in the immediate vicinity of d, which appear in a large number of LOCI-reduced image. By applying the same characterization method to these planet/speckle candidates, we verify that these candidates can be safely rejected, as they are either detected in as single roll (with a minimum SNR threshold of 3), or in one instance (red circle) the speckle detected in both rolls but with positions incompatible with error bars (see Figure \ref{Fig:FalseAlarm}). Note that this particular speckle also appears in Figure 1, directly above the planet d in roll 2.}\label{5speckles}
\end{figure}

\begin{figure}[htbp]
\center
\resizebox{0.5\hsize}{!}{\includegraphics{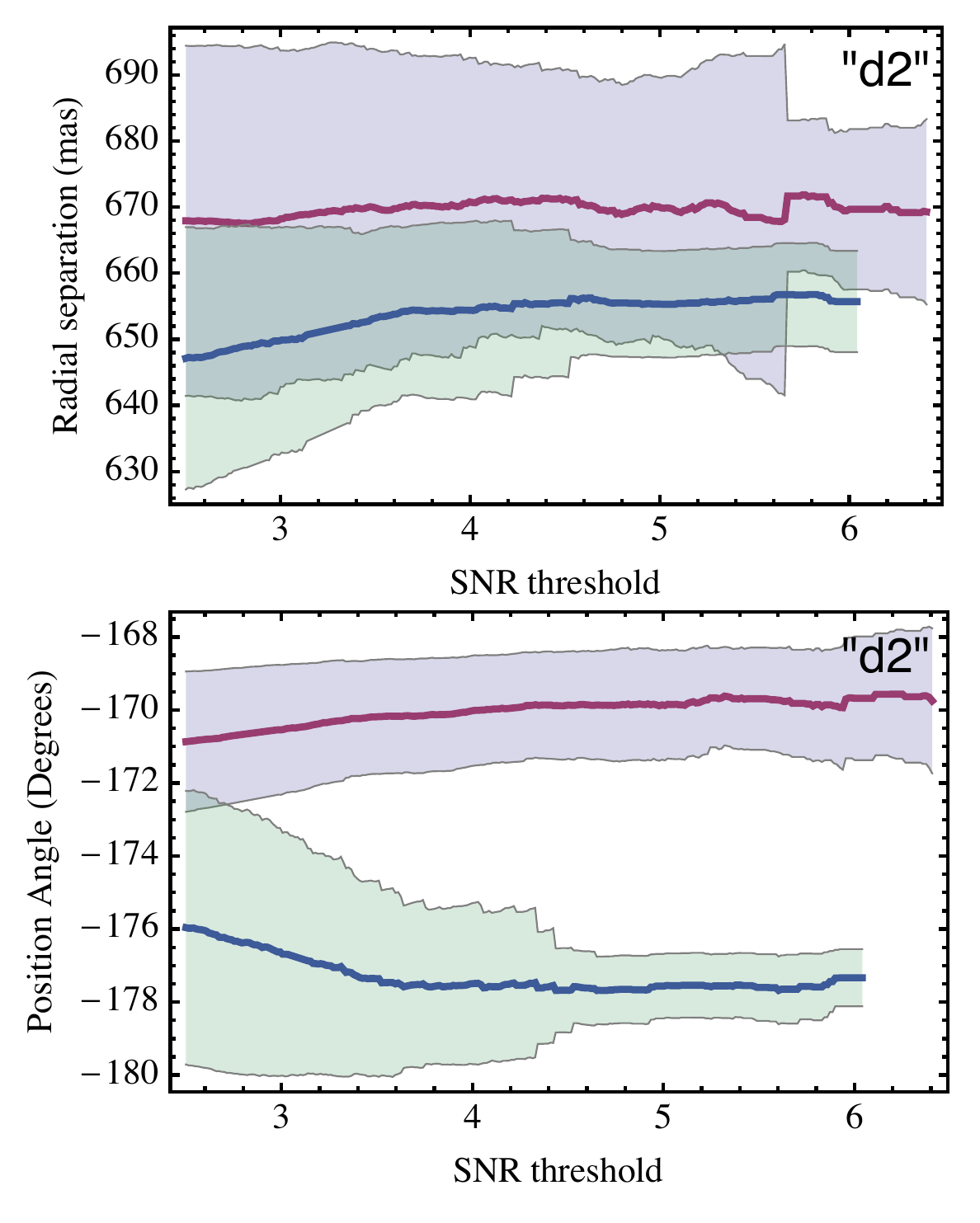}}
 \caption{Astrometric study for one of the five planet/speckle candidates identified in Figure \ref{5speckles} showing that this candidate can be safely rejected. As the SNR increases, the measured astrometry from roll to roll does not converge as it does for measured astrometry for the real planet d, and it is not consistent between the two rolls given the error bars. Furthermore, the study we perform on astrometric biases using fake planet injections show that the potential error in measurement is significantly smaller than this roll to roll discrepancy. We note that the other four speckles studied in Figure \ref{5speckles} were only detected in a single roll.}\label{Fig:FalseAlarm}
\end{figure}

The first test is to select the five brightest speckles that appear in the vicinity of d (Figure \ref{5speckles}) in many of the LOCI-reduced images and consider them as planet d candidates. Starting from the position estimates for these five planet/speckle candidates, we analyze the statistics of the astrometry in a three-pixel diameter area centered around the estimated position as a function of SNR threshold, exactly as we did for the real planets. 
All five candidates are very conservatively rejected. Four of them are only detected in a single roll with a minimum SNR threshold of 3. In one instance the candidate is detected in both rolls but the two measured positions are not consistent with the error bars (Figure \ref{Fig:FalseAlarm}), which indicates that it must be a speckle. 
We note that for Figure \ref{Fig:FalseAlarm} and Figure \ref{Fig:SNRthreshold}, the error bars show the true spread of measured astrometric positions (i.e. one sigma). This means that for the rejected candidate, there were no LOCI images within one sigma for which the astrometry matched in both rolls among the 6,300 tested images.

\begin{figure}[htbp]
\center
\resizebox{0.6\hsize}{!}{\includegraphics{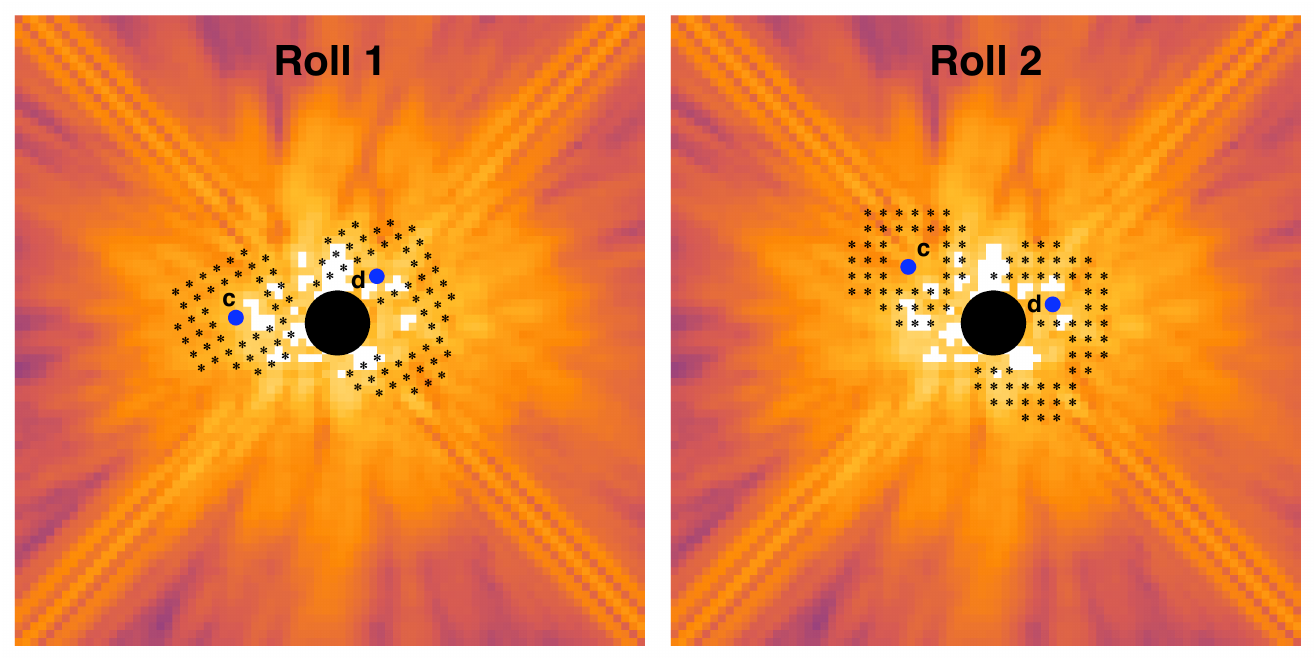}}
 \caption{The 96 positions studied for estimating the false alarm probability. Each pixel position in the figure to the right ( HR 8799 at first orientation) has a corresponding position in the figure to left ( HR 8799 second orientation), rotated 29.9 degrees about the star center. We note again that in the entire reduced data set for HR 8799 has 12,600 reduced LOCI images; because we use a single-pixel LOCI method, we do not reduce the entire 80x80 image because of how expensive it is computationally and instead focus on regions around the planet (Figure \ref{fig1}). Thus, we select positions to test for the false alarm probability based on radial separation and the amount of reduced image space available. The two small circles in each image show the positions of planets c and d, and the small empty region below planet d is where we looked for the undetected planet e. The large circle in the center shows the location of the coronagraphic hole. We exclude these regions for the false alarm analysis. }\label{Fig:FalseAlarmGrid}
\end{figure}

The second test is to estimate the probability of false alarm, which we define as the probability to detect two speckles in both rolls at the correct rotation angle, and within the astrometric error bars. 
For that we defined 96 pairs of pixels positions separated by 29.9 degrees between both rolls, as illustrated in Figure \ref{Fig:FalseAlarmGrid}. For each of these pairs of positions, we apply the complete astrometric analysis for all the 12,600 images. Note that the pixel positions avoid the position of c and d planet and the empty area corresponds to the region where we searched for planet e. 
For five pairs of pixel positions a speckle is detected in both rolls at an SNR level higher than 3. However, in two cases the astrometric measurements are not consistent between each roll and thus we have a total of three qualifying false alarms, which yields an estimate for the probability of false alarm of $\sim3\%$. 
This number is conservative in that for real planets we also observe an improvement of the astrometry with increasing SNR, whereas in these three false alarm cases we observe little to no convergence or improvement as the SNR increases. 

The third test is to study the photometry of the detected planets b, c and d.
First we calibrated the effect of LOCI on the photometry for each planet  (even with masking, LOCI will tend to subtract some fraction of the planet PSF [see section 3.2.4]). We injected fake planets in the reference PSF from the library at the true planet position in both rolls, and measured the post-LOCI photometry using the best 50 parameter sets for each planet and computed the LOCI equivalent throughput as a function of planet brightness. We generated a calibration curve between true photometry and post-LOCI measured photometry. Finally we measured the planet photometry in the HR 8799 system for the best 50 images for each planet, and estimated the true photometry as the mean of these measurements and the photometric error from their standard deviation. The results are presented in Table \ref{tablePhotometry}.

\begin{table*}
\begin{center}
\caption{Photometric measurements.\label{tablePhotometry}}
\begin{tabular}{cccccc}
\tableline\tableline
Planet & \multicolumn{2}{c}{Count Rate (DN s$^{-1}$)} & \multicolumn{3}{c}{Vega Magnitude} \\
\tableline
  & Roll 1 & Roll 2 & Roll 1 & Roll 2 & Mean\\
\tableline
 HR 8799b & $24.16 \pm 2.54$ & $15.53 \pm 2.42$ & $18.36 \pm 0.11$ & $18.85 \pm 0.17$ & $18.61 \pm 0.14$\\
 HR 8799c & $53.99 \pm 8.29$ & $28.66 \pm 2.80$ & $17.50 \pm 0.18$ & $18.18 \pm 0.11$ & $17.84 \pm 0.14$\\
 HR 8799d & $71.18 \pm 34.78$ & $63.23 \pm 16.54$ & $17.30 \pm 0.50$ & $ 17.35 \pm 0.31$ & $17.33 \pm 0.40$\\
\tableline
\end{tabular}
\tablecomments{F160W Photometry of the three planets in the HST data. For Comparison the photometry obtained by \citet{LMD09} for HR 8799b is $18.54\pm 0.12$, which is consistent with our measurement.}
\end{center}
\end{table*}

The photometry for HR 8799b agrees with the results from \citet{LMD09} (18.54 $\pm$ 0.12). Assuming identical colors F160W-H for all three planets, we extrapolate the photometry for planet c and d from the H-band photometry from \citet{MMB08}, and find that planet c ought to be $\sim$17.6 mag and d should be $\sim$17.5 mag. These values are in excellent agreement with the photometry we present in Table \ref{tablePhotometry} apart from the value for planet c in Roll 2, where the planet falls close to the diffraction spikes resulting in an over subtraction of the planet PSF. 
The consistency of the photometry for planet d between both rolls and with the expected value based on the H-band photometry reinforces the confidence level in the detection of the true planet.

\section{Orbits and analysis}

Several dynamical studies of this planetary system based on a three-planet system have shown that stable solutions for this system are limited to small regions of the phase space involving mean motion resonances (MMR) that stabilize the orbits over time scales comparable to the age of the star. Apart from these few islands of stability the system appears chaotic over timescales shorter than the age of the system. The most important resonance for the system's stability is 1d:2c, with also the possibility of a double resonance 1d:2c:4b. Other solutions have been identified based on the initial astrometric data published by \citet{MMB08} involving a 1d:1c solution where both planets have similar eccentricities \citep{GM09}. With the recent discovery of the fourth planet \citep{MZK10} and then confirmed by dynamical studies will certainly have to be revisited, but we assume that stable solutions with four planets are a subset of the stable solutions for three planets. Several publications \citep{RKS09,LMD09,MRS10,WCD11} suggest a moderate inclination for the system ($13-40\deg$), from considerations on stellar rotational velocity, disk inclination, and stability. Based on the dynamics for a face-on coplanar system \citet{MHC10} find that systems with moderate eccentricities ($e\sim 0.08-0.2$) are disrupted over short time scales ($10^4-10^5$ yr). With a fourth planet very close to d \citep{MZK10} it is likely that this will push the constraints to lower eccentricities for d. 

We combine our astrometric measurements for the three planets b, c and d using the 1998 HST data with the previous astrometric measurements for these planets in 2002 ( Table 1 in \citet{FIT09}), 2004, 2007 and 2008 ( Table 1 in \citet{MMB08}), 2007 ( Table 1 in \citet{MMZ09}), and 2008, 2009 ( Table 2 in \citet{CBI11}) to study the possible orbits and discuss stability based on previous dynamical studies for this system \citep{GM09,RKS09,FM10,MHC10,MZK10}. 
The results presented in the following section use all of these astrometric data. We note that our orbit-fitting results are a strong function of which data is included, and furthermore are highly dependent upon the error bars given for these astrometric data points. We find large variations in the orbit-fitting results when removing different sets of data points; this can indicate the presence of uncalibrated systematic errors in the published estimates, in particular that could be due to absolute North calibration errors between different telescopes. For this reason, we apply more conservative rejection levels to the orbit-fitting solutions than customary.

We implemented the method described by \citet{cat10} to fit Keplerian orbits to these data and explored orbit fitting for each planet independently first, and then under the hypothesis of MMRs. MMRs add very strong constraints to the orbits fit because all planet periods and semi-major axes are no longer free parameters.  A single period and semi-major axis are fitted for one of the planets in resonance. The other periods are directly obtained from the resonance definition, and the other semi-major axes from Kepler's third law since the mass of the star is unique. In this case, the fitting method by \citet{cat10} needs to be modified.
The mass of the star can either be considered as a known quantity in the fit of $1.47\pm0.3$ $M_{\sun}$\citep{GK99}, or as as free parameter to obtain a dynamical mass measurement under the hypothesis of the MMR. We also use a weighted least-square method to account for different error bars between observations.
 Although MMRs do not require exact integer period ratios for the osculating orbits, this is a reasonable approximation given our astrometric uncertainties and limited time baseline, and we first consider the case of exact integer period ratios between the planets. The goal of this study is to check whether the stable solutions identified by previous dynamical studies with three planets are still compatible with the HST data points, within error bars. 
In particular, we do not prove that our best-fit solutions are stable. Since there are now four known planets around HR 8799, new dynamical studies will be necessary to investigate the full set of stable solutions that are compatible with the ensemble of astrometric data.

\begin{figure}[htbp]
\center
\resizebox{0.6\hsize}{!}{\includegraphics{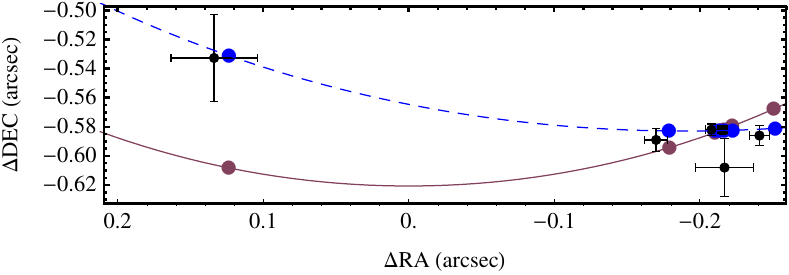}}
\caption{  Top: Orbit fit for planet d in the case of a face-on system with or without eccentricity. We marginally reject the case of a face-on circular orbit for planet d ($\chi^2=21.8$ with corresponding p-value of 0.03).
Based solely on planet d astrometry, the $\chi^2$ values favor eccentricity and/or inclination, which is consistent with predictions purely based on dynamical constraints.  
}\label{Fig:OrbitdAlone}
\end{figure}

 For b and c planets, the orbital motion over the $\sim 10$ years baseline is not sufficient to place any constrain on inclination or eccentricity based on independent orbit fitting for each planet. For planet d, however, the fit strongly favors an eccentric and/or inclined orbit. Indeed, a face-on circular orbit for d  is unlikely, with $\chi^2=21.8$ with { 11} degrees of freedom  (p-value=0.03). The best fit for a face-on orbit is obtained for a high eccentricity ($\epsilon_d = 0.63$), with { $\chi^2=8.76$} and 10 degrees of freedom (p-value=0.56). The best fit for a circular orbit (searching between 0-90 degree inclination) is obtained with a high inclination of 48.1 degrees, with  $\chi^2=8.75$ and 11 degrees of freedom (p-value=0.64). We show these two extreme cases solutions for a face-on system (with and without eccentricity) in Figure \ref{Fig:OrbitdAlone}.  
 \citet{BBJ11} performed a similar study of the eccentricity and inclination of planet d using previously published astrometric data, with approximately 2-year baseline. We note that our results for the eccentricity and inclination of planet d are in good agreement with their results. Furthermore with a 10-year baseline, we strengthen their conclusion that planet d must either be slightly eccentric or inclined. For a face-on circular orbit they find the inclination to be greater than 43 degrees within a $99.73\%$ confidence limit, with a best fit of 63 degrees (compared to our 48.1 degrees). For a face-on eccentric orbit, they find planet d must have eccentricity  $\geq0.4$ within a $99.73\%$ confidence limit, with a best fit eccentricity of $\sim 0.7$, compared to our eccentricity $e_d=0.63$ in the face-on configuration.

We studied three possible MMRs discussed in the literature: 1d:2c, 1d:2c:4b, and 1d:1c using a combined $\chi^2$ for the three planets orbits. 
All dynamical studies with three planets agreed that the 1d:2c resonance plays a major role in stabilizing this system, with the possibility of a double resonance 1d:2c:4b as well. 
We also investigate the constraints our new data points place on 1d:1c, which seems more unlikely given the eccentricities required ($\sim$ 0.25) and the close proximity with the recently discovered planet e. 

 We limit our study to the case of co-planar systems and explore a range of inclination from 0 to 60 degrees to cover a larger range of values than discussed in the literature. 
For 1d:2c and 1d:2c:4b we assume circular orbits for b and c, and only explore eccentricity for d. For 1d:1c we assume both eccentricities to be equal for simplicity.
We find that the temporal baseline is not large enough to properly constrain stellar mass when assuming the 1d:1c and 1d:2c resonances. We then use the published stellar mass of 1.47 $M_{\sun}$ \citep{GK99}. When assuming the 1d:2c:4b resonance it is possible to fit a dynamical mass and the best fit corresponds to star mass of  1.56 $M_{\sun}$, which is consistent with the errors bars from \citet{GK99}. 

\begin{figure}[htbp]
\center
\resizebox{0.6\hsize}{!}{\includegraphics{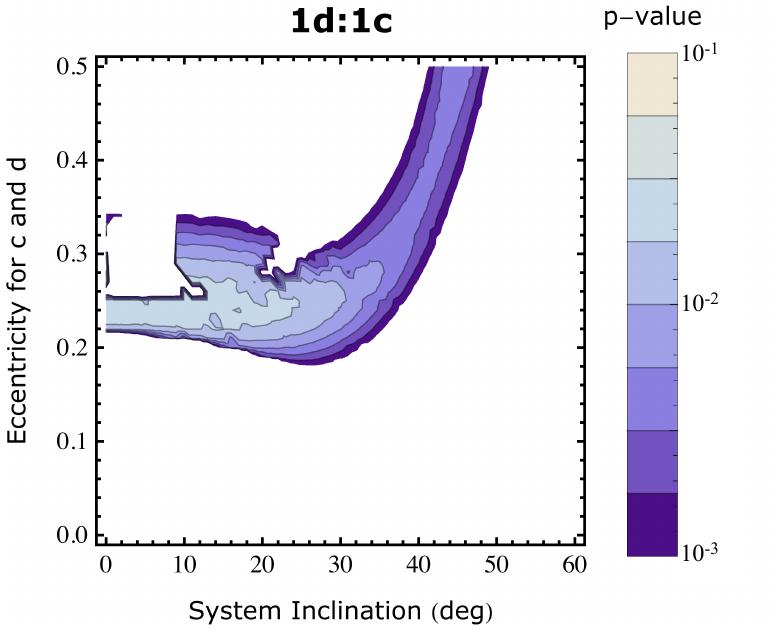}}
\caption{ P-values from the chi-square distribution of the fitted orbits in the case of the 1d:1c main mean motion resonance that correspond to stable solutions. We assume identical eccentricity for c and d. The vast majority of this paramter space is completely rejected, and the best fit region has a p-value of $\sim 0.03$. We note the stable solution by \citet{GM09} for 1d:1c has a p-value of 0.03.}\label{Fig:pValues11}
\end{figure}

\begin{figure}[htbp]
\center
\resizebox{0.6\hsize}{!}{\includegraphics{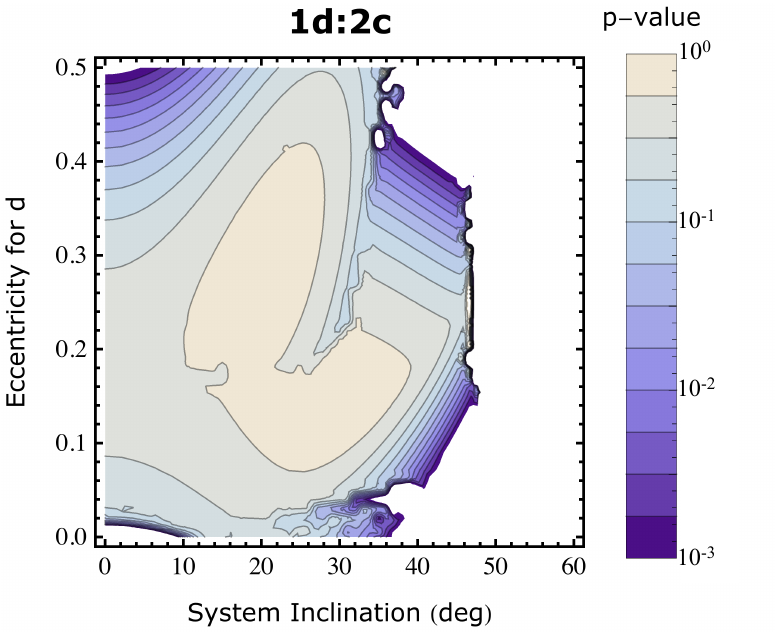}}
\caption{{ P-values from the chi-square distribution of the fitted orbits in the case of the 1d:2c main mean motion resonance that correspond to stable solutions. We note that most of the parameter space is compatible, with the exception of higher inclinations ($i>49$ degrees). We also note the rejection of a face-on circular solution for this configuration, and highly eccentric d in face-on system.}}\label{Fig:pValues12}
\end{figure}

\begin{figure}[htbp]
\center
\resizebox{0.6\hsize}{!}{\includegraphics{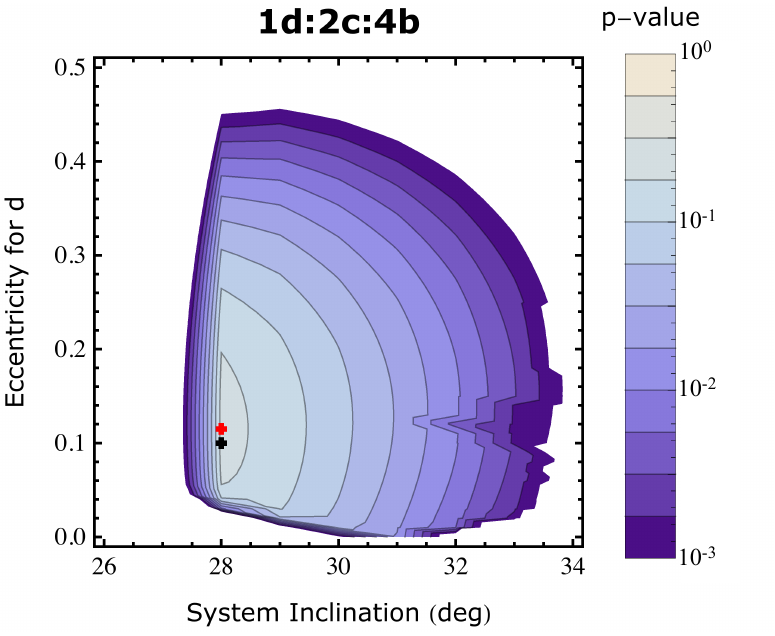}}
\caption{{ P-values from the chi-square distribution of the fitted orbits in the case of the 1d:2c:4b main mean motion resonance that correspond to stable solutions for a star mass of 1.47 $M_\sun$. We assume circular orbits except for d. For 1d:2c:4b the stable solution identified by \citet{GM09} based on the discovery data \citep{MMB08} has very similar eccentricity to our best fit solution ($e_d=0.115$, marked by the red cross), and remains compatible with the new data set, albeit for a larger inclination. The red cross marks the best solution we find when we fit for the star mass (best fit corresponds to star mass of 1.56 $M_{\sun}$ [see Figure \ref{Fig:OrbitsMMR}])}}\label{Fig:pValues124}
\end{figure}

 We provide p-value maps in Figures \ref{Fig:pValues11}, \ref{Fig:pValues12}, and \ref{Fig:pValues124} for the range of inclinations and eccentricities discussed above for 1d:1c, { 1d:2c,} and 1d:2c:4b MMRs. Assuming a known star mass (i.e. not fitting the star mass), we have  27 degrees of freedom for 1d:1c and  28 for 1d:2c, and  50 degrees of freedom for 1d:2c:4b, which we also verified using Monte-Carlo simulations to construct the empirical $\chi^2$ distributions for each of the cases we tested.  

{ The p-value map for 1d:1c (Figure \ref{Fig:pValues11}) shows that most of the parameter space for eccentricity and inclination is completely rejected (i.e. the white region, which corresponds to p-values lower than 0.001) assuming both planets have equal eccentricities for simplicity. Less parameter space would probably be ruled out if we removed this assumption, but this would add another dimension to the problem. 
 The best fit stable solution identified by \citet{GM09} for the 1d:1c MMR does not have exactly identical eccentricities for both planets. We tested the fit for this particular configuration with eccentricities 0.267 for d, and 0.248 for c, and with an inclination of 11.4 deg. For this specific solution we find $\chi^2= { 42}$ with { 27}  degrees of freedom. This corresponds to a p-value of  ${0.03}$ and therefore this MMR solution is unlikely. The best fit region for the 1d:1c resonance assuming identical eccentricities for simplicity also has a p-value of $\sim 0.03$. Now that a fourth planet has been discovered this solution seems more unlikely to be stable because of the close proximity with e and because of the eccentricities involved with c and d. Dynamical studies and more astrometric measurements will be needed to confirm this conclusion.} 

{ For the 1d:2c (Figure \ref{Fig:pValues12}) resonance we can rule out the case of a circular face-on coplanar system (p-value $\simeq0$), as well as systems with inclinations greater than 40 to 50 deg depending on the eccentricity of d.
This calculation assumes that the periods follow exact integer ratios (strict 1:2 resonance), i.e. that the semi-major axes $a_c/a_d$ are in a perfect ratio as well $(P_c/P_d)^{2/3}=2^{2/3}$ according to Kepler's law. 
Under this assumption of a strict 1d:2c resonance we need to invoke eccentricity for d and/or inclination of the system. We note that our best fit region for this parameter space has a p-value of  $\sim 0.7$.} 

{ The p-value map for 1d:2c:4b (Figure \ref{Fig:pValues124}) places much stronger constraints on the possible solutions that are compatible with the data, since the period and semi major axes for all three planets depend on two parameters only (one period and the star mass suffice to determine the tree semi-major axes, assuming a strict resonance). Imposing stric ratios between the semi-major axes is a very strong constraint because it is equivalent to setting the projected separation into a specific geometry. 
Assuming a star mass of 1.47 $M_{\sun}$ we obtained a well-identified best-fit solution with inclination and eccentricity $i= { 28.0}$ deg and $e_d= { 0.115}$.}
At a very conservative $0.1\%$ p-value rejection level there are no solutions for the { 1d:2c:4b resonance} compatible with the data for eccentricity $e_d> { 0.46}$, and for inclinations {$i<27.3$ and $i>33.9$} (see Figure \ref{Fig:pValues124}).
The particular case of a face-on system with all circular orbits is strongly rejected with a p-value $\sim 0$ ($\chi^2\simeq 7000$) under this 1d:2c:4b hypothesis. 

\begin{figure*}[htbp]
\center
\resizebox{0.9\hsize}{!}{\includegraphics{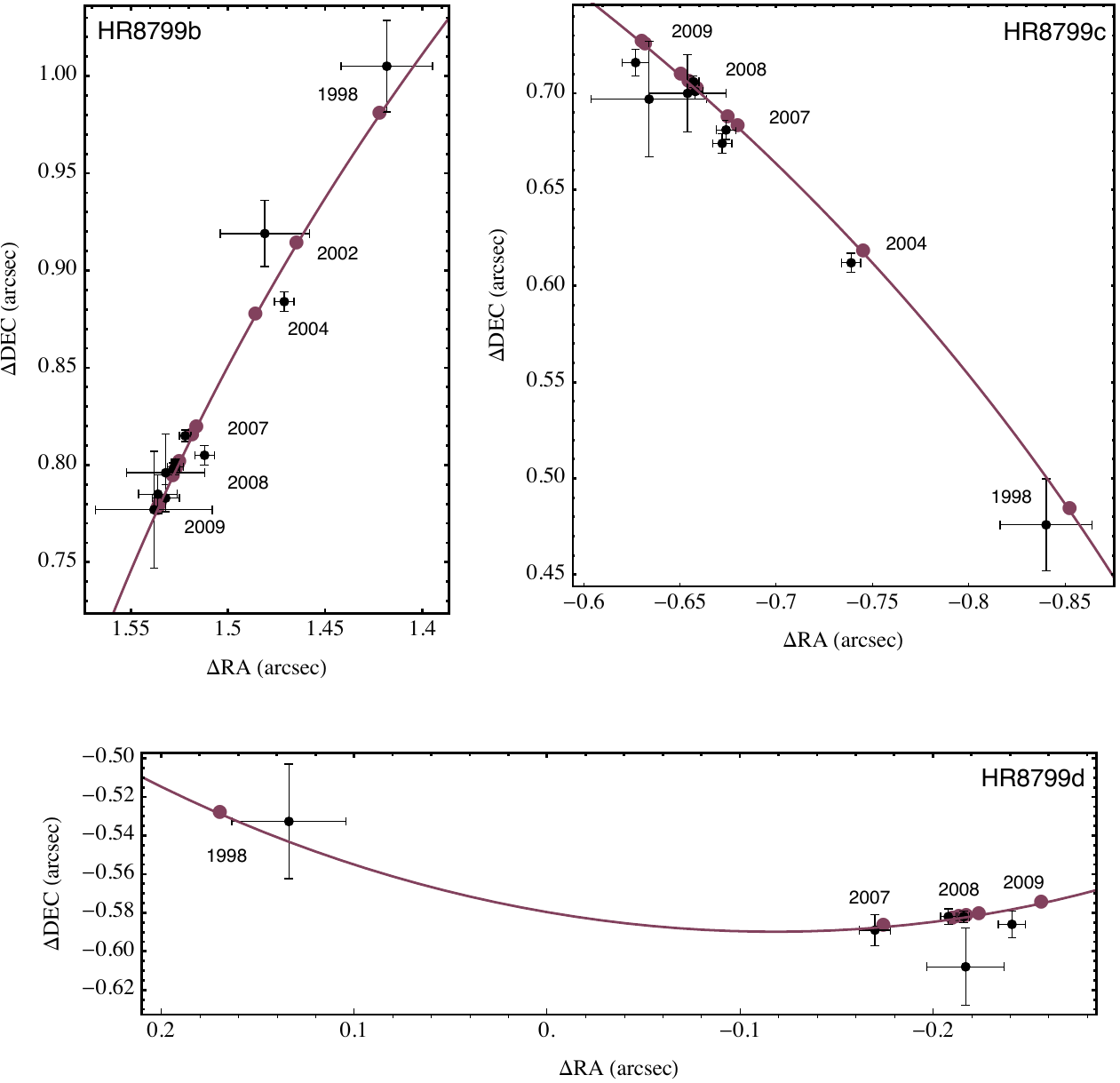}}
\caption{Orbit fits for planets b, c, and d based on the the best fit solution for the 1d:2c:4b mean motion resonance, assuming a coplanar system with circular orbit for b and c, and eccentric orbit for d. The p-value of the overall solution is { 0.26} and the star mass is { 1.56} $M_\sun$ }\label{Fig:OrbitsMMR}
\end{figure*}

With a star mass of 1.47$\pm0.30$ $M_{\sun}$ the corresponding p-value for the best fit solution is  { 0.23} for 1d:2c:4b. { If the mass of the star is also included in the fit parameters, we find a marginally higher p-value of 0.26 for a star mass of 1.56 $M_{\sun}$ (compatible with the error bars on the mass), with eccentricity $e_d= 0.100 $ and inclination $i= 28.0$ deg. It is interesting to note that our best-fit eccentricity for d is very close to the 1d:2c:4b best fit solution by \citet{GM09}.} We show the corresponding orbits for this solution in Figure \ref{Fig:OrbitsMMR}, and we summarize the orbital parameters for these two solutions in Table \ref{tableorbits}.

\begin{figure}[htbp]
\center
\resizebox{0.5\hsize}{!}{\includegraphics{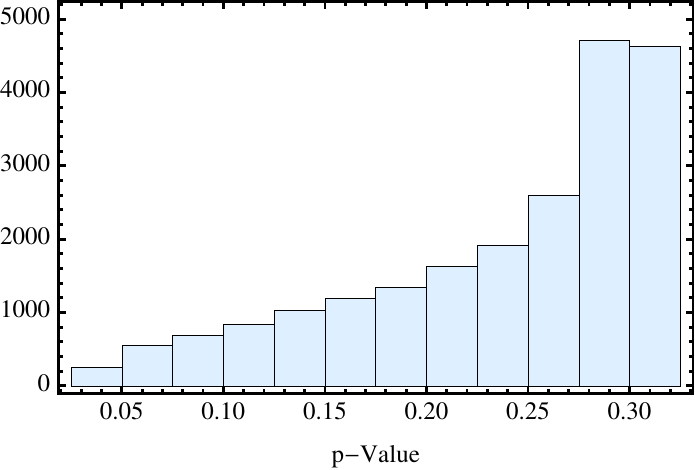}}
\caption{{ Histogram of p-values for $\sim20000$ realizations where we vary the the periods of planets b, c, and d  around the values for our best fit solution  for the 1d:2c:4b resonance, and assuming a uniform spread for the periods of $2\%$ of the period of d planet. This shows that although our fit assumed exact integer ratios, our conclusions remain compatible with the data if we consider small departures from the exact integer period ratios for the osculating orbits, as it can be expected in a MMR configuration.}}\label{Fig:RandomMMR}
\end{figure}

{ Since we have not integrated these orbit solutions over the lifetime of the system, we have not tested if our best fit solutions are dynamically stable. 
Instead our approach verifies that the stable 1d:2c:4b solution identified by \citet{GM09} with 0.075 eccentricity for d, very low eccentricities for b and c (0.008 and 0.012) and a star mass of 1.455 $M_\sun$ remains compatible with the data (p-value= { 0.18}), if we assume our best fit inclination of  { 28.0} deg. 
Because of short-term perturbations in orbital motions due to planet-planet interactions, the periods of the osculating orbits may not follow exact integer ratios and present librations around the exact resonance values. We used a Monte-Carlo simulation to test the impact of small departures from a strict resonance. In Figure \ref{Fig:RandomMMR} we show the histogram of the p-values of the orbit fit for 20,000 realizations where the periods of each planet were randomized around our best fit solution by an amount corresponding to $2\%$ of the period of planet d, assuming a coplanar system with inclination $i= 28.0$ deg, circular orbits for b and c and eccentricity for d $e_d= 0.1 $. This shows that our conclusion remains compatible with the data even in the presence of non-exact integer period ratios for the osculating orbits. A complete dynamical study will remain necessary to conclude on the stability of this resonance, for example following the method described by \citet{FM10} for the three-planet system before the discovery of the fourth planet.}

\begin{table*}
\begin{center}
\caption{Summary of the best fit solutions for MMR 1b:2c:4b\label{tableorbits}}
\begin{tabular}{cccccccccc}
\tableline\tableline 
\tableline
 Star mass ($M_{\sun}$)  & Planet & Period (yr) & inclination (deg) & $\omega$ & eccentricity & $\Omega$ &  sma (AU) & p-value & T0  \\
\tableline
1.47		&	d	&	115.9	&	28.0	& 75.28 &	0.115 &35.9	& 27.0 & 0.23 & 2005.3\\
1.47		&	c	&	231.7	&	28.0	& NA &	0.0 &	35.9&42.9  & 0.23 & 1997.5\\
1.47		&	b	&	463.3	&	28.0	& NA &	0.0 &	35.9&68.1  & 0.23 & 2043.1\\
\hline
1.56		&	d	&	112.4	&	28.0	& 80.2 &	0.1 &35.5	& 27.0 & 0.26 & 2003.7\\
1.56		&	c	&	224.9	&	28.0	& NA &	0.0 &	35.5& 42.9 & 0.26 & 1936.7\\
1.56		&	b	&	449.7	&	28.0	& NA &	0.0 &	35.5& 68.0 & 0.26 & 2047.1\\		

\tableline
\end{tabular}
\tablecomments{This table summarizes the main orbital parameters for the best fit solutions assuming a 1d:2c:4b resonance with a mass of 1.47$\pm0.30$ $M_{\sun}$ \citep{GK99}, or 1.56 $M_{\sun}$ (our best-fit dynamical mass under this resonance assumption). 
}
\end{center}
\end{table*}

\section{Conclusion}
In this paper we studied the HST NICMOS coronagraph archival data set of HR 8799 from 1998, using the LOCI PSF subtraction algorithm.
We improved previous results by \citet{LMD09} by optimizing the LOCI algorithm and detecting three planets (b, c and d) in these data. The fourth planet was not detected in this data set.
Our LOCI reduction takes advantage of the quality improvements of the LAPL PSF library \citep{SSS10}, which includes better darks, flats, and bad pixels calibration in addition to a large number of reference PSFs (we use 466 references PSFs to reduce the HR 8799 data).

LOCI enables extremely efficient PSF subtraction: the detection sensitivity we obtain is improved by one order of magnitude compared to a classical PSF subtraction (roll-deconvolution) with this data set. 
The LOCI subtraction can therefore slightly affect the planet PSF shape, which can in turn generate sub-pixel astrometric errors. 
Some LOCI implementations tend to significantly modify the throughput and shape of the PSF and introduce astrometric biases up to 2/3 of a pixel ($\simeq 50$ mas), as measured using artificial planets injected in one of the reference PSFs. In order to overcome this potential problem, we studied a number of variants of the LOCI algorithm to determine the least biased configuration using fake injected planets in multiple reference PSFs. 
The best solution we found for this dataset uses the masking technique introduced by \citep{MMV10} for a single pixel LOCI method, with an exclusion zone equivalent in size to a planet PSF size, which is excluded from each optimization region in the algorithm. 

The second cause of astrometric errors is due to residual speckle noise, which also affects the astrometry at the sub-pixel level. 
To overcome this second problem we developed a method for astrometric measurements on LOCI reduced images. We vary the algorithm parameters to introduce some speckle diversity in the final images by generating a large number of LOCI-reduced images (12,600). For each LOCI-reduced image, we use matched-filtering to determine the planets position and SNR. We then studied the statistics of the astrometry as a function of SNR, and used simulations with fake planets. Since we do not find a single set of LOCI parameters that provides the highest SNR everywhere in the image, our approach to explore a large parameter space guarantees good images for each position of interest. This approach also provides a large number of astrometric measurements, and therefore enable the determination of statistically significant error bars and the evaluation of potential biases using Monte-Carlo simulations. 

Because the detection of d in these HST data is very challenging, we studied the possibility of a false alarm in great detail. We conservatively estimate the probability of false alarm to be of the order of $3\%$, which stems from the joint false-positive detections in both rolls purely based on astrometric agreement. In addition, our detection is strengthened because the photometry of our detected d planet is consistent between both rolls. 
We also rejected the possibility that one of the five brightest speckles in the same vicinity may in fact be the real planet d. 

We derive astrometric positions for b, c and d for this epoch, and find that our measurement for b is consistent with prior results by \citet{LMD09}. 
We then fit Keplerian orbits for each planet using all available published data for b,c and d. 
In light of recent dynamical studies for this system \citep{GM09,RKS09,FM10,MHC10,MZK10}, we select a few interesting cases of stable configurations and study their compatibility with the additional data points from HST, which provide a ten year baseline with the discovery image from 2008. 

{ For the individual orbits of planets b and c the data does not place strong constraints on possible orbits, except that we can marginally reject a face-on circular orbit for planet d. Our data for planet d favors either a face-on system with high eccentricity, a highly inclined circular orbit, or an inclined and moderately eccentric orbit. These results are consistent with \citep{BBJ11}. A high eccentricity will likely be ruled out with the addition of the fourth planet by future dynamical studies.}

{ In the case of mean motion resonances, we are able to place strong constraints on possible orbits compatible with the data. Following \citet{MZK10} we assume that the stability of the four-planet system will likely be a subset of the stable solutions with three planets. Even with three planets, all dynamical studies have shown that the stability of the system is limited to a few mean motion resonances, mainly dominated by the interaction between c and d. 

{ The stable solution involving a 1d:1c resonance from \citet{GM09} is unlikely with the addition of the HST data. In any case, it is likely that the presence of a fourth planet will make this configuration unstable given the significant eccentricity involved and the close proximity for c, d, and e.

For the 1d:2c MMR, we can rule out the possibility of a system with high inclination and also a circular coplanar face-on system. The rest of the parameter space remains mostly compatible with the data.

We find that the stable solution for the double resonance 1d:2c:4b identified by \citet{GM09} remains compatible with the new data, with the particularly interesting result that our best-fit is very close to this stable solution. 
This double resonance imposes very strong physical constraints on the fit (one period and the star mass suffice to define the three periods and the three semi-major axes). 
We can thus rule out most of the parameter space of inclinations and eccentricities assuming circular orbits for b and c. The completely circular face-on system hypothesis is rejected under the 1d:2c:4b hypothesis, and possible inclinations for a coplanar system are confined to a small range around 28.0 deg ($27.3-33.9$). The best-fit solution is obtained for a low eccentricity for d ($e_d=0.10$) assuming circular orbits for b and c. The range of possible eccentricities compatible with the data is moderate, and we can very conservatively rule out eccentricities larger than 0.46. Although our best fit p-values are not very high, this can be explained by indications of possible astrometric biases between datasets with different telscopes, presumably due to systematics such as absolute north calibration. The fact that our best-fit eccentricity for d corresponds closely to the stable solution, and that our best-fit solution is robust to small departures from the exact integer period ratios suggests that dynamical simulations should be carried out to test the 1d:2c:4b resonance hypothesis further with all data available to date.

With the recent discovery of a fourth planet \citep{MZK10}, the already challenging dynamics of this system becomes even more interesting.
These new astrometric data points from HST archival data will help future dynamical studies improve the understanding of this system, without having to wait too long for planets to move along their orbits. { This work also provides new photometric information for c and d taken in the F160W filter.}
Since the fourth planet e has a short enough period ($\sim 50$ yr) its orbit should be constrained relatively quickly. 
This study underlines the value of the HST archive for direct imaging of exoplanets and the importance of such archives and PSF libraries for future missions such as the James Webb Space Telescope.}}

\acknowledgements 
The authors thank Glenn Schneider and Dean Hines for discussions and help providing us with the re-calibrated NICMOS data from AR 11279. 
The authors also thank Neill Reid, Bruce Macintosh, David Lafreni\`ere, Marshall Perrin, and Jay Anderson for help and discussions. JBH thanks Ben Sugerman for recommending working on this project. 
This work was performed in part under contract with the California Institute of Technology (Caltech) funded by NASA through the Sagan Fellowship Program and also supported by the STScI Director's Discretionary Research Fund.


\bibliography{/Users/soummer/Documents/Work/Library/Library/MasterBiblio} 
\bibliographystyle{apj}

\end{document}